# From Mott Insulators to Quantum Metals

S. Er-Rahmany[1,2], M. Loulidi[2], A. El Kenz[2], A. Benyoussef[2,3], M. Azzouz[1,4*]

1. International University in Rabat, Aerospace Engineering School, LERMA, Sala El Jadida, Morocco
2. Mohammed V University in Rabat, Faculty of Science, Condensed Matter Laboratory and Interdisciplinary Sciences (LaMCScI), Rabat, Morocco
3. Hassan II Academy of Sciences and Techniques, Rabat, Morocco
4. Department of Physics, Laurentian University, Ramsey Lake Road, Sudbury, Ontario, Canada P3E 2C6

*Corresponding author. Email: mohamed.azzouz@uir.ac.ma; mazzouz@laurentian.ca

**Abstract:**
High critical temperature cuprate superconducting materials are composed of copper oxide layers and interlayer charge reservoirs. When not doped, these cuprates are antiferromagnetic insulators. We propose to design new materials by combining alternating layers of parents of hole-doped and electron-doped of these cuprates and modifications thereof. Our goal is to find undoped cuprates that can be either an antiferromagnetic insulator or a quantum metal. The term quantum metal means a metal characterized by long range antiferromagnetic order or only strong antiferromagnetic correlations, i.e., it is thus a stable ground state against any other perturbations. The new metallic states sought here could be precursors to new superconducting states in the absence or presence of doping. Using the density functional theory, we report on two compounds $\{La\}\{Pr\}CuO_4$ and $\{La\}\{V\}CuO_4$ that illustrate the different physics described above. The curly brackets mean that the preparation of these compounds shall be done by depositing a layer containing Pr, then one $CuO_2$ layer, then finally the La layer in $\{La\}\{Pr\}CuO_4$ for example. The configuration formed by the positions of the charge reservoir atoms with respect to the $CuO_2$ layer is an important factor in the new procedure we propose here. This paper reports on the X-ray diffraction, electronic, optical, and magnetic properties of these hypothetical materials. We found that $\{La\}\{Pr\}CuO_4$ is a Mott insulator, but $\{La\}\{V\}CuO_4$ is an undoped correlated quantum metal with long-range order. Our calculations were performed using the linearized plane wave method (FP-LAPW) implemented within the Wien2k software.

## I. Introduction

More than thirty years elapsed since the discovery of high-temperature superconductivity by Bednorz and Müller [1] in $La_{2-x}Ba_xCuO_4$ with critical temperature $T_C = 30$ K [2]. But the mechanism of high-temperature superconductivity has yet to be discovered. This mechanism differs from Bardeen-Cooper-Schieffer's (BCS) theory, which deals with phonon-mediated superconductivity in low-temperature conventional superconductors [3]. Meanwhile, the search for new superconducting materials did not stall at least experimentally; currently innovative materials like $H_2S$ [4], graphene bilayer [5], $LaH_{10}$, $CaC_6$ and $YbC_6$ [6], etc., have been investigated.

Indeed, the quest for innovative materials with critical superconducting temperatures higher than those already realized is a challenging task. Recently, some progress has been achieved. Drosdov et al. realized conventional superconductivity in H$_2$S below ~203 K under very high-pressure conditions, [4]. Also, conventional superconductivity was reported in lanthanum hydride LaH$_{10}$ below ~250 K at a pressure of 170 GPa [7]. Unconventional superconductivity has been discovered in the purely two-dimensional strongly-correlated bilayer graphene, with very low critical temperature 1.7 K, but with a very high ratio of this temperature and Fermi energy given that the Fermi surface consists of very small pockets in this material [8]. Amid all this, the high-$T_C$ cuprates continue to hold the record highest critical temperature in atmospheric pressure conditions. Indeed, to the best of our knowledge, Hg$_{0.8}$Tl$_{0.2}$Ba$_2$Ca$_2$Cu$_3$O$_{8+\delta}$ holds the record highest $T_C = 135$ K at one atmosphere [9].

In high-$T_C$ cuprate superconductors (HTSC), the interplay between antiferromagnetic (AF) order or strong AF correlations and superconductivity is a challenging and still open issue. It is believed that the understanding of this issue will very likely lead to understanding the mechanism of high-$T_C$ superconductivity. It is in this spirit that, in this work, we propose an approach for searching for innovative materials based on HTSCs and having a richer interplay between antiferromagnetism and metallicity at or near zero doping. We begin with undoped parent compounds of HTSCs and address the question of changing the physical properties of these materials from a Mott insulator to a quantum metal by changing chemical constituents and proposing a new way for preparing these hypothetical materials. The quantum metallic state proposed here is characterized by either long-range AF order or strong AF correlations.

This approach has been motivated as follows. The phase diagram of HTSCs is characterized by two important phases in addition to the superconducting phase, namely the AF phase near zero doping, and the pseudogap (PG) phase, which borders the superconducting phase and which occurs for doping levels slightly greater than those of the AF region. The rotating antiferromagnetism theory (RAFT) describes the PG phase in terms of the rotating antiferromagnetic (RAF) hidden order. This approach gives rise to a phase diagram that is comparable to the experimental one, and is based on competing orders, i.e., the competition between superconductivity and RAF order, [10]–[15]. Note that RAF order can be interpreted as a remainder of real three-dimensional (3D) AF order that occurs in the vicinity of zero doping. When the coupling responsible for RAF order is not considered in RAFT, the phase diagram consists of only superconductivity, which is optimal near zero doping. When this coupling is considered back in RAFT, superconductivity is destroyed in the vicinity of zero doping. The following idea, born from this analysis, motivated the present work: if a mechanism that enhances superconductivity near zero doping at the expense of AF order and PG behavior is found then higher superconducting critical temperatures could perhaps be realized. Our approach consists of proposing the use of chemical substitution in layered high-$T_C$ cuprates and a special material's preparation in order to engineer new compounds with modified and enhanced properties: we propose materials made of alternating layers of electron-doped and hole-doped cuprates and investigate the effect of different configurations when the positions of reservoir's atoms are exchanged between upper reservoir layers and lower ones with respect of the CuO$_2$ layer. In short, the present work can be considered as the first step in the quest for innovative high-$T_C$ materials where superconductivity could be enhanced at the expense of antiferromagnetism and/or AF correlations in the immediate vicinity of zero doping. Here, we focus on undoped materials only to illustrate our new approach. We use the density functional theory (DFT) to investigate the effect of different configurations of atoms in the charge reservoirs within the unit cell on the physical properties like optical conductivity, AF order, and density of states (DOS).

In this paper, we report on the proposal of combining the compounds La$_2$CuO$_4$ and Pr$_2$CuO$_4$ to obtain a new material hereafter labeled as {La}{Pr}CuO$_4$. Then we substitute Pr by Vanadium, V to get {La}{V}CuO$_4$. The curly brackets mean that in the proposed material, the positions of atoms

La and Pr in the unit cell matter. Two configurations are shown in Figure 1. We use the DFT to study the effect of changing positions on the local AF magnetization and DOS. Any changes of magnetization and of the DOS in the vicinity of the Fermi energy will be a signal for a potentially interesting change in the properties of the material. For atoms' substitution, we replaced Pr by vanadium V (Figure 2) which belongs in the same column in the periodic table.

In Section II, the computation methodology is outlined. Section III is dedicated to the results and discussion. These results concern the new hypothetical materials {La}{Pr}CuO$_4$ and {La}{V}CuO$_4$ but also include the compound La$_2$CuO$_4$. The electronic structure, DOS, X-ray diffraction profiles, charge density, and optical conductivity are calculated and discussed. Conclusions are drawn in Section IV.

## II.     Materials and Methodology

The computations were performed using the code Wien2k [16], which is an implementation of the Full Potential-Linearized Augmented Plane Wave (FP-LAPW) [16], [17], in the framework of the DFT [18], [19]. We used the generalized gradient approximation (GGA) parameterized by Perdew, Berke and Erenzehof [20], [21] for the exchange correlation potential, because it is known to produce more accurate band structures, contrary to the local density approximation (LDA) [22]. It, for example, yields a lower energy compared to the LDA approximation. We begin by recalculating the electronic, magnetic, and optical properties of the material La$_2$CuO$_4$ using Wien2k. We used 500 **k**-points and $Rk_{max} = 7$ which converge well the total energy of the system. To account for the energy correlations of the highly localized 3d copper orbitals, we used the DFT+$U$ within the GGA+$U$ approximation [23]. In this method, the onsite (copper's d orbital) interaction is described by a pair of parameters: The Coulomb repulsion $U$ and exchange interaction $J$ [24]. The values $U = 4$ eV and $J = 0.4$ eV were considered in our calculation for copper [25]. We used a 1×1×1 cell of the antiferromagnetic orthorhombic crystal structure containing 28 atoms. This cell is sufficient because it contains 4 copper atoms with 2 in each AF sublattice.

For the new hypothetical materials {La}{Pr}CuO$_4$ (Figure 1) and {La}{V}CuO$_4$ (Figure 2), we used the GGA+$U$ approximation with $U = 7$ eV [26], $U = 4$ eV [25], and $U = 5$ eV [27] for orbitals 4f of Pr, 3d of Cu, and 3d of V, respectively. We also used $J = 0.8$ eV [26], $J = 0.4$ eV [25], and $J = 1$ eV [27] for orbitals 4f of Pr, 3d of Cu, and 3d of V, respectively. We used the XCrySDen code to visualize all structures [28], and the VESTA code to calculate the X-ray diffraction pattern for {La}{Pr}CuO$_4$ and {La}{V}CuO$_4$ [29].

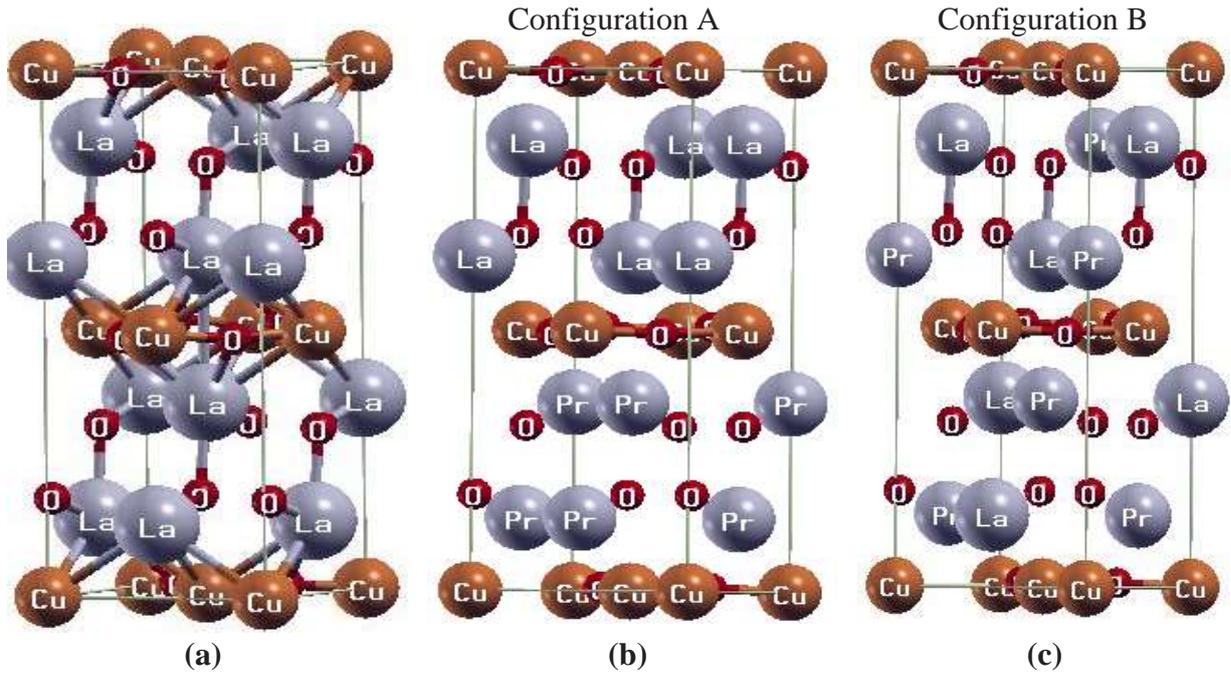

**Figure 1.** The crystalline structure of (a) La$_2$CuO$_4$, (b) Configuration A for {La}{Pr}CuO$_4$ without mixing of La and Pr in the charge reservoir, and (c) Configuration B for {La}{Pr}CuO$_4$ with mixing between La and Pr.

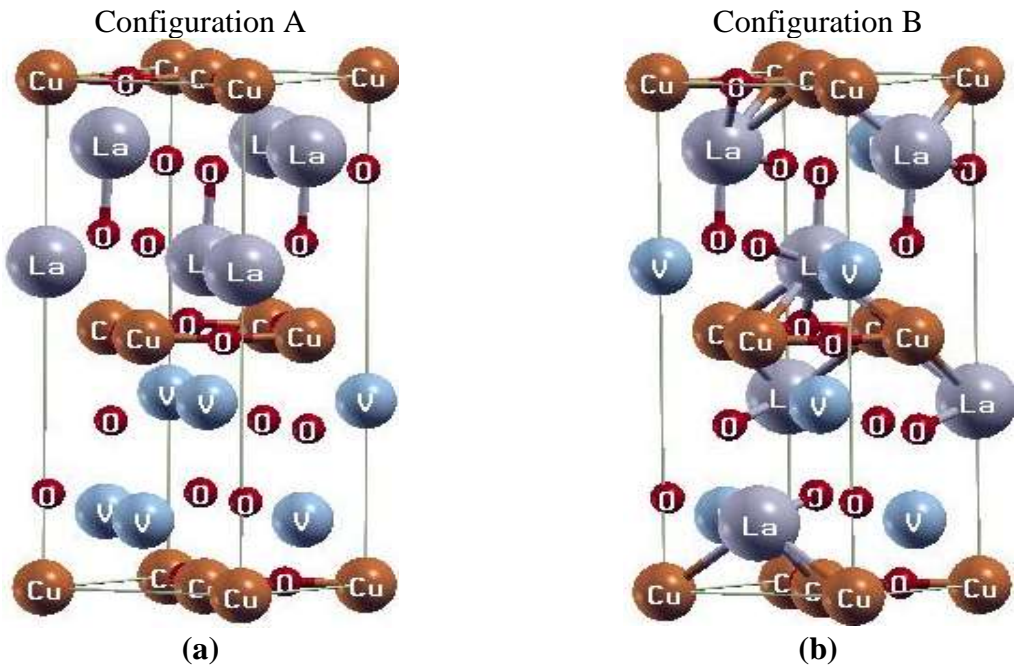

**Figure 2.** The crystalline structure of (a) {La}{V}CuO$_4$ without mixing of La and V in the charge reservoir, (b) {La}{V}CuO$_4$ with mixing between La and V.

### III. Results and discussion

#### 1. The compound La$_2$CuO$_4$
#### 1.1. Band structure

The La$_2$CuO$_4$ compound crystallizes in the low-temperature orthorhombic structure (Figure 1a) with Bmab space group (N°. 64) [30], and has been studied by Czyżyk et al. [31]. The atomic positions of La$_2$CuO$_4$ are summarized in Table 1.

**Table 1.** Atomic positions of $La_2CuO_4$ [30].

| Atoms | Multiplicity and Wyckoff | x | y | z |
|---|---|---|---|---|
| La | 8f | 0 | -0.00672 | 0.36145 |
| Cu | 4a | 0 | 0 | 0 |
| $O_1$ | 8e | 1/4 | 1/4 | -0.00672 |
| $O_2$ | 8f | 0 | 0.3406 | 0.18367 |

To determine the ground-state properties, we optimized the crystal structure of the system using the Birch-Murnaghan Equation of State (BM-EOS) [32]. The procedure is to allow for the relaxation of positions of atoms in the system. Once equilibrium is reached, we calculated the physical properties of interest. The optimized lattice parameters ($a$, $b$, and $c$) and angles ($\alpha$, $\beta$, and $\gamma$) of the unit cell of $La_2CuO_4$ are shown in Table 2. The results are in good agreement with recent literature [30], [33], [34].

**Table 2.** The optimized parameters of $La_2CuO_4$.

| Parameter and angle | $a$ (Å) | $b$ (Å) | $c$ (Å) | $\alpha$ | $\beta$ | $\gamma$ |
|---|---|---|---|---|---|---|
| Experimental [30] | 5.357 | 5.406 | 13.143 | 90° | 90° | 90° |
| Experimental [33] | 5.335 | 5.420 | 13.106 | ---- | ---- | ---- |
| GGA [34] | 5.352 | 5.576 | 13.101 | ---- | ---- | ---- |
| GGA | 5.358 | 5.483 | 13.135 | 90° | 90° | 90° |
| GGA+U | 5.360 | 5.410 | 13.139 | 90° | 90° | 90° |

Figure 3 shows the X-ray diffraction (XRD) intensity profile for material $La_2CuO_4$ calculated by the software VESTA [29]. The peak with highest intensity occurs at (020) for $2\theta \approx 30°$. The XRD profile calculated here is in agreement with experimental data [35].

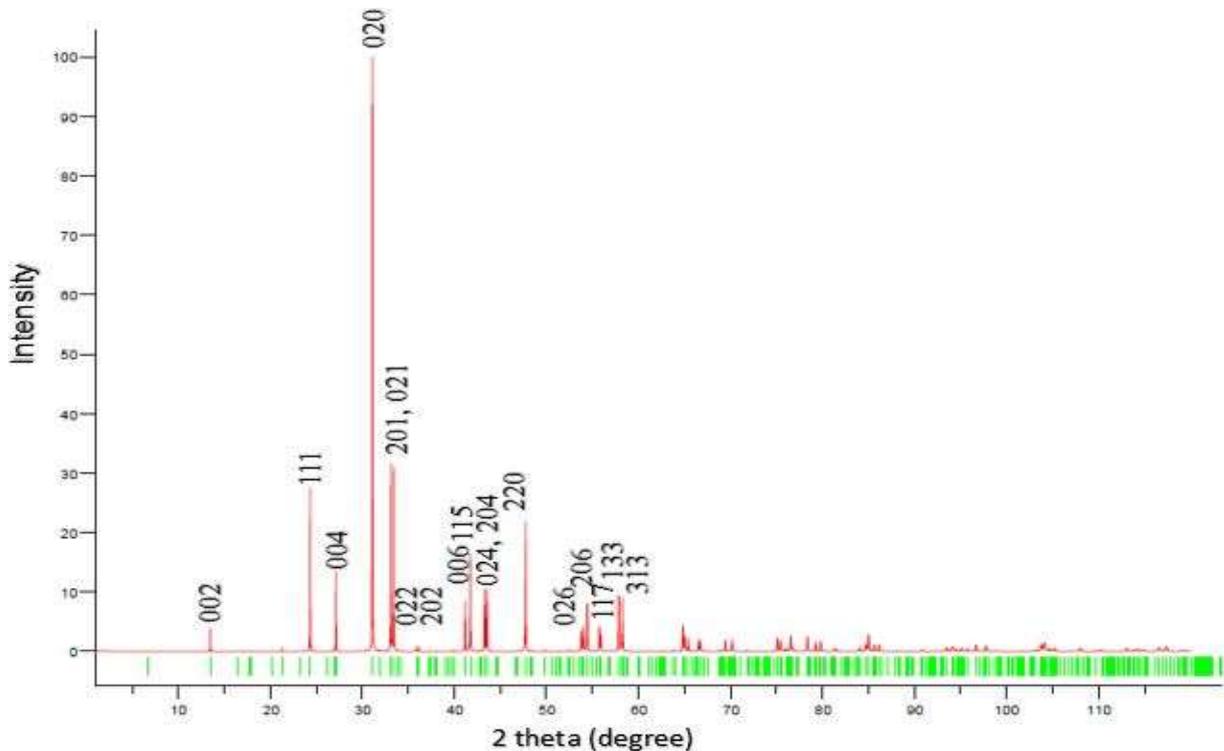

**Figure 3.** The XRD profile for $La_2CuO_4$ calculated by VESTA.

In our calculation, $[Xe]5d^16s^2$, $[Ar]3d^{10}4s^1$ and $1s^22s^22p^4$ states were considered as electronic configurations for La, Cu, and O, respectively. The valance of La, Cu and O are $3^+$, $2^+$ and $2^-$, respectively. Using the GGA approximation, we recalculated the total DOS and band structure of La$_2$CuO$_4$ and displayed the results in Figure 4. The calculation was performed using the polarized spins' feature of Wien2k because the AF order is known to occur in this compound. This AF order is confirmed by the GGA+$U$ calculations (see below). The DOS in Figure 4a indicates that this compound has a nonmagnetic character because the total magnetic moment is practically zero; $0.01\mu_B$ (Table. 3). In the absence of local Coulomb repulsion on copper's atoms, a metallic behavior, as is well known, characterizes the band structure shown in Figure 4b. Physically, the DFT predicts a metallic state since the bands are not fully filled (the band formed by copper's d electrons is half filled).

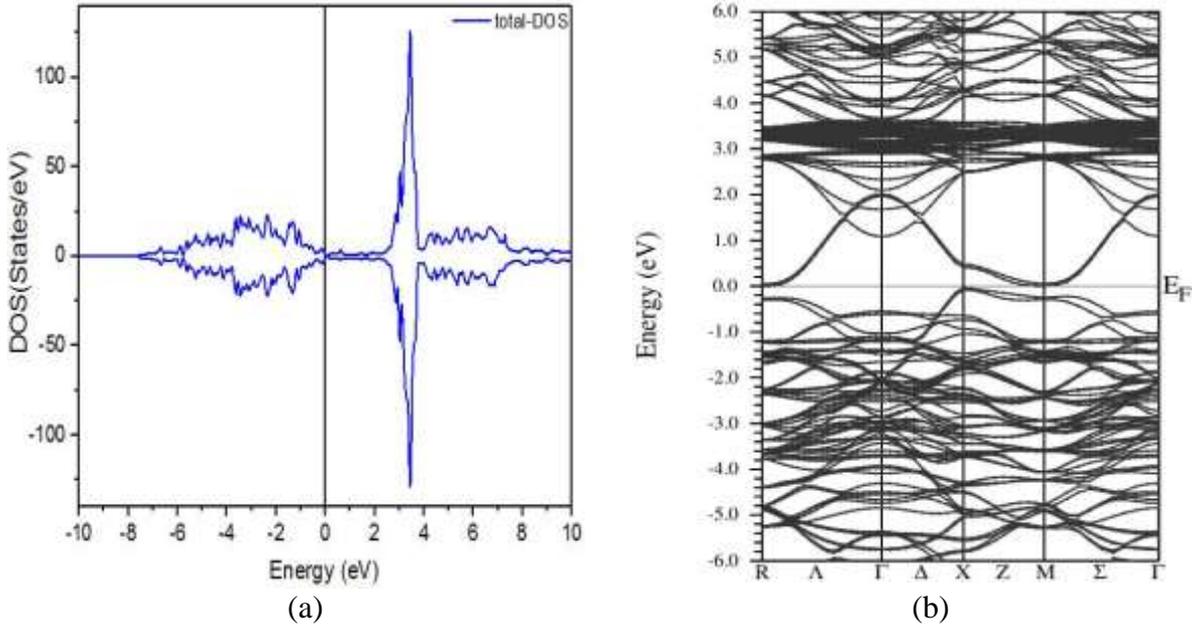

(a) (b)

**Figure 4.** (a) The total DOS, and (b) band structure of La$_2$CuO$_4$ calculated within the GGA approximation are shown. The Fermi energy is taken as the origin for energies.

According to band theory, La$_2$CuO$_4$ should thus be a metal, however it was reported to be insulating experimentally because of the strong onsite Coulomb repulsion [36]. For this reason, GGA+$U$ has been used with $U = 4$ eV for copper's 3d orbitals. We note from Figures 5a and 6 that the total DOS and band structure have greatly changed due to the inclusion of the Hubbard $U$ term. In Figure 6, a gap of 1.5 eV results from electron correlations due to Coulomb repulsion. These results are in good agreement with recent literature [37] where the (linear-spin-density approximation) LSDA+$U$ approximation was used [38]. The gap we find is indirect since the minimum of the conduction band (CB) and the maximum of the valence band (VB) are shifted in agreement with results in literature [34], [39].

**Table 3.** The partial magnetic moments, total magnetic moment, and gap of La$_2$CuO$_4$.

| Approximation | | | GGA | GGA+U | Experimental | Ref. [31] | Ref. [37] | Ref. [40] |
|---|---|---|---|---|---|---|---|---|
| Magnetic moment ($\mu_B$) | | La | 0.0013 | 0.0008 | ------- | ------- | ------- | ------- |
| | Cu | Cu1 | 0.26 | 0.58 | 0.5 [41], [42] | 0.62 | ------- | 0.495 |
| | | Cu2 | -0.26 | -0.58 | | | | |
| Total magnetic moment ($\mu_B$) | | | 0.01 | 0 | ------- | ------- | ------- | ------- |
| Gap (eV) | | | 0 | 1.5 | 2.00 [43] | 1.65 | 2.00 | 1.00 |

A recent study [40] showed that an improved treatment of the first principles of the antiferromagnetic ground state of La2CuO4 can be obtained by the strongly-constrained-and-appropriately-normed (SCAN) meta-GGA [44] within the DFT framework, without using the Hubbard $U$ parameter. The SCAN meta-GGA yielded a better estimate of the band gaps compared to GGA+$U$ approximation [45].

In the case of GGA approximation (where the polarized spins' feature of Wien2k is used even in the absence of the Hubbard term $U$), the magnetic moment of copper atoms is $0.26\mu_B$ (Table 3). This value is close to the literature value $0.2\mu_B$ in Ref. [34]. When the Hubbard term is added, the compound becomes an AF insulator with a staggered magnetic moment on copper atoms equal to $0.58\mu_B$, which is comparable to the experimental value $0.5\mu_B$ [41], [42]. This value is consistent with $1\mu_B$ for $S = 1/2$ when the Landé factor is $g = 2$, [46].

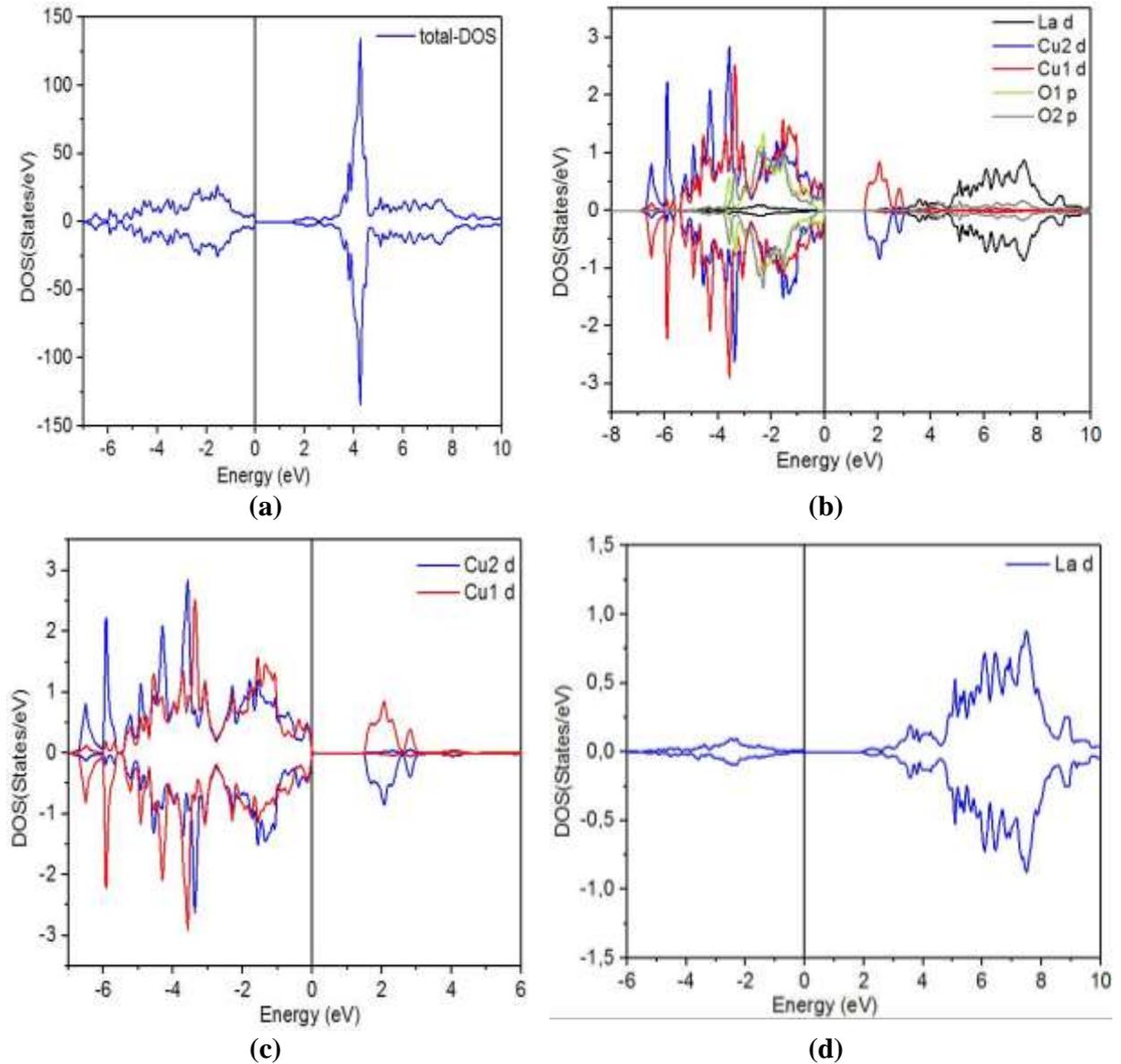

**Figure 5.** (a) Total DOS of $La_2CuO_4$, (b) Partial DOS of orbital d of La, orbital d of Cu, and orbital p of O calculated within the approximation GGA+$U$, (c) Partial DOS of spin-up and spin-down, Cu1 and Cu2, and (d) Partial DOS of orbital d of La.

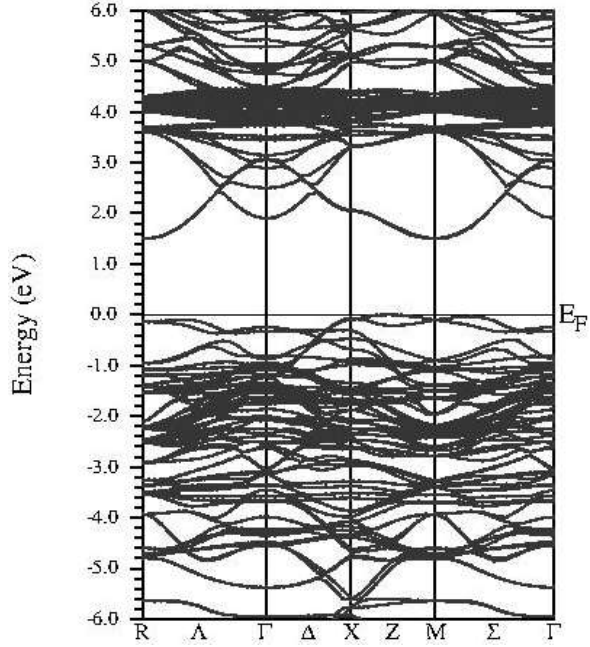

**Figure 6.** The band structure of La$_2$CuO$_4$ calculated within the GGA+$U$ approximation.

### 1.2. Optical conductivity for La$_2$CuO$_4$

Figure 7a shows the evolution of the real part of the optical conductivity from 0 to 14 eV along the $x$ (in *ab* plane) and $z$ (*c*) directions of La$_2$CuO$_4$ within GGA+$U$. The real part of the optical conductivity is given by, [47],

$$\sigma(\omega) = -\frac{\omega}{4\pi}\text{Im}\varepsilon_{ij}(\omega), \quad (1)$$

with $\varepsilon_{ij}$ being the dielectric function, and ω the frequency. Figure 6a shows clearly that the optical conductivity decreases rapidly for energies higher than ~8.5 eV, whereas it increases for energies between 0 and 8.5 eV. This overall behavior compares well to existing experimental data [48], [49]. In Figure 5a, the total DOS is highly peaked at ~5 eV, then it is peaked again at ~8.5 eV where the optical conductivity has also the maximum value. Also, in Figure 7a, characteristic features at $E_1$= 1.01 eV, $E_2$ = 1.5 eV, $E_3$ = 1.8 eV, and $E_4$ = 2.3 eV are displayed. The peaks at $E_2$, $E_3$ and $E_4$ are due to direct transitions in the directions $M$, R and Γ from the VB to CB, respectively. $E_1$ corresponds to the optical gap, which is smaller than the electronic gap (Kohn-Sham gap) because the GGA approximation neglects electron-hole (exciton) contributions [50]. The latter corrects the gap and can be addressed within the GW approximation [51] and time dependent DFT calculation, which are beyond the scope of the present paper [52], [53]. The optical gap has been estimated using Tauc's equation, [54],

$$(\alpha h\nu)^n = A(h\nu - E_g), \quad (2)$$

where $E_g$ is the optical gap, $\alpha$ the absorption coefficient, $h\nu$ the photon energy, $A$ a material-dependent constant, and $n = 2$ for a direct transition or 1/2 for an indirect transition. In our case, the material La$_2$CuO$_4$ is characterized by an indirect bandgap; therefore $n = 1/2$. Fitting the *ab* absorption coefficient using equation (2) near the optical gap yields the curve shown in Figure 7b, with $E_g = 1.01$ eV. The latter is in good agreement with experiment [48].

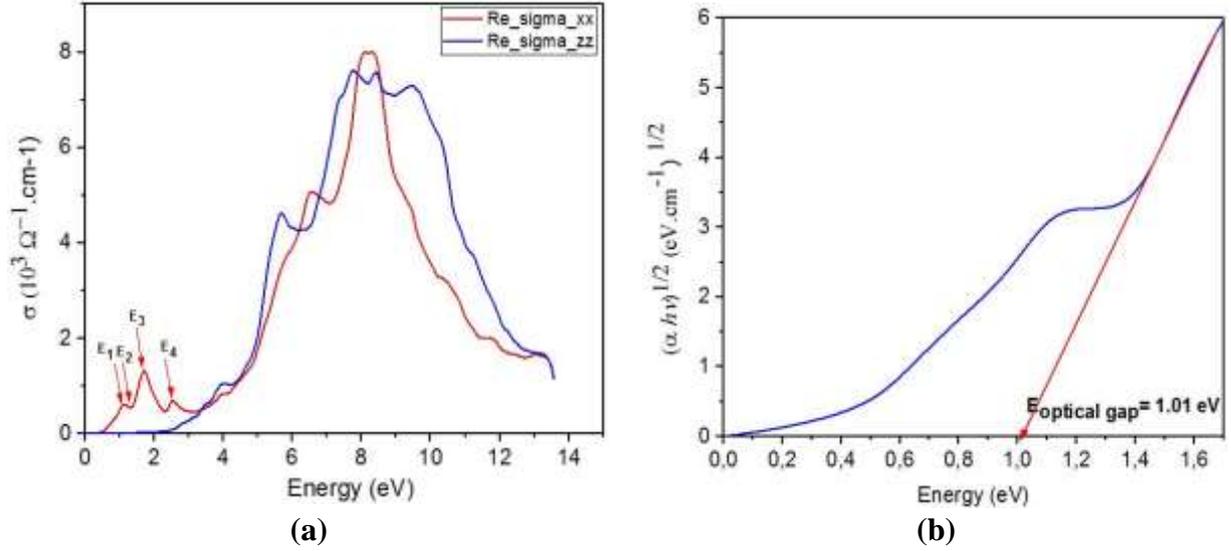
**Figure 7.** (a) Optical conductivity of $La_2CuO_4$, and (b) optical gap.

## 2. The hypothetical compound $\{La\}\{Pr\}CuO_4$

Kuberkar et al. investigated the structural and superconducting properties of $La_{1-x}PrSr_xCuO_4$ with $0.1 \leq x \leq 0.25$, [55]. They found from resistivity versus temperature measurements that the metallicity of the compound increases with Sr doping $x$. On the other hand, they found that the superconducting transition temperature increases with $x$, reaching a maximum at $x = 0.2$ where $T_C = 25$ K, then decreases for $x > 0.2$. Here we propose our new materials, i.e., we substitute atoms by others and propose a new way for growing materials. The proposed material $\{La\}\{Pr\}CuO_4$ must be grown by depositing layer by layer, the bottom charge reservoir layer, $CuO_2$ layer, and upper charge reservoir layer. Here, we substitute one La atom by Pr in $La_2CuO_4$. First, we checked that $\{La\}\{Pr\}CuO_4$ is stable chemically in the two configurations A and B by calculating the formation energies:

$$E_{\text{Formation}} (A) = E_{\text{system}} - E_{Pr} - E_{La} - E_{Cu} - 4E_O$$
$$= -62726.5680 \text{ Ry},$$

$$E_{\text{Formation}} (B) = E_{\text{system}} - E_{Pr} - E_{La} - E_{Cu} - 4E_O$$
$$= -62726.5684 \text{ Ry}.$$

These energies are negative and large in absolute value, and these compounds are stoichiometrically neutral since Pr and La atoms have the same degree of oxidation, namely $La^{3+}$ and $Pr^{3+}$.

### 2.1. X-ray diffraction profile

Figure 8 shows the (XRD) intensity profiles for the two $\{La\}\{Pr\}CuO_4$ structure configurations calculated using the software VESTA [29]. Note that the two configurations have the same intensity profile because cations $La^{3+}$ (radius 1.160 Å) and $Pr^{3+}$ (1.126 Å) have comparable radii so that it does not matter significantly where these ions are located [56].

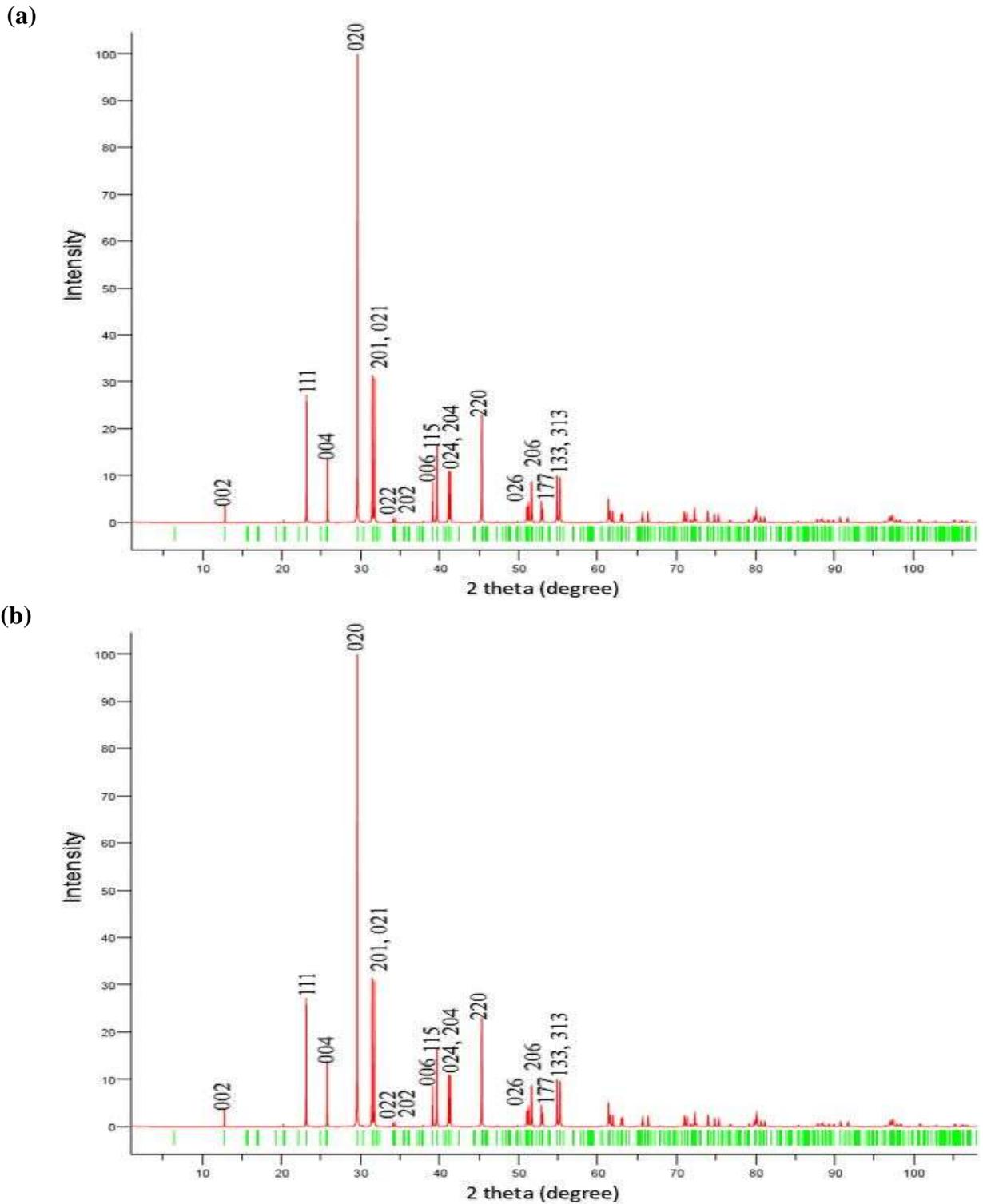

**Figure 8.** The XRD profile calculated by VESTA [29] for (a) {La}{Pr}CuO$_4$ without mixing of La and Pr in the charge reservoir, and (b) {La}{Pr}CuO$_4$ with mixing of La and Pr.

The parameters of the elementary cell ($a$, $b$, and $c$) and angles ($\alpha$, $\beta$, and $\gamma$) for the two {La}{Pr}CuO$_4$ structure configurations are shown in Table 4.

**Table 4.** The optimized lattice parameters of {La}{Pr}CuO$_4$.

| Lattice parameters and angles | $a$ (Å) | $b$ (Å) | $c$ (Å) | $\alpha$ | $\beta$ | $\gamma$ |
| --- | --- | --- | --- | --- | --- | --- |
| GGA+U Configuration A | 5.551 | 5.603 | 13.610 | 90° | 90° | 90° |
| GGA+U Configuration B | 5.550 | 5.606 | 13.611 | 90° | 90° | 90° |

## 2.2. Band structure and DOS

We calculated the electronic structure and DOS corresponding to configurations A and B using GGA+$U$ approximation and compared the results with those of $La_2CuO_4$. Figures 9 and 10 display the total DOS and partial DOS including those of the d electrons on copper atoms Cu1 of sublattice 1, and Cu2 of sublattice 2. Sublattices 1 and 2 correspond to local up and down AF ordered moments. Note that the results of the two {La}{Pr}$CuO_4$ configurations are comparable with those of $La_2CuO_4$ especially near the Fermi energy (the top of the VB). In Figures 9b and 10b, it found that the orbital f of atom Pr has a higher density compared to that of other atoms (Cu, La, and O). This density occurs, however, at energies far away from the Fermi energy.

The band structures in Figure 11 corresponding to configurations A and B display almost the same value for the gap (Table 5). Note that the gap in {La}{Pr}$CuO_4$ is smaller than in $La_2CuO_4$. The substitution of half of Pr atoms with La atoms in $La_2CuO_4$ decreases the gap and increases slightly the magnetic moment of copper (Table 5). The hypothetical material remains thus a Mott insulator. The smaller values of the gap could lend the hypothetical materials {La}{Pr}$CuO_4$ to optoelectronic applications in the vicinity of the visible spectrum.

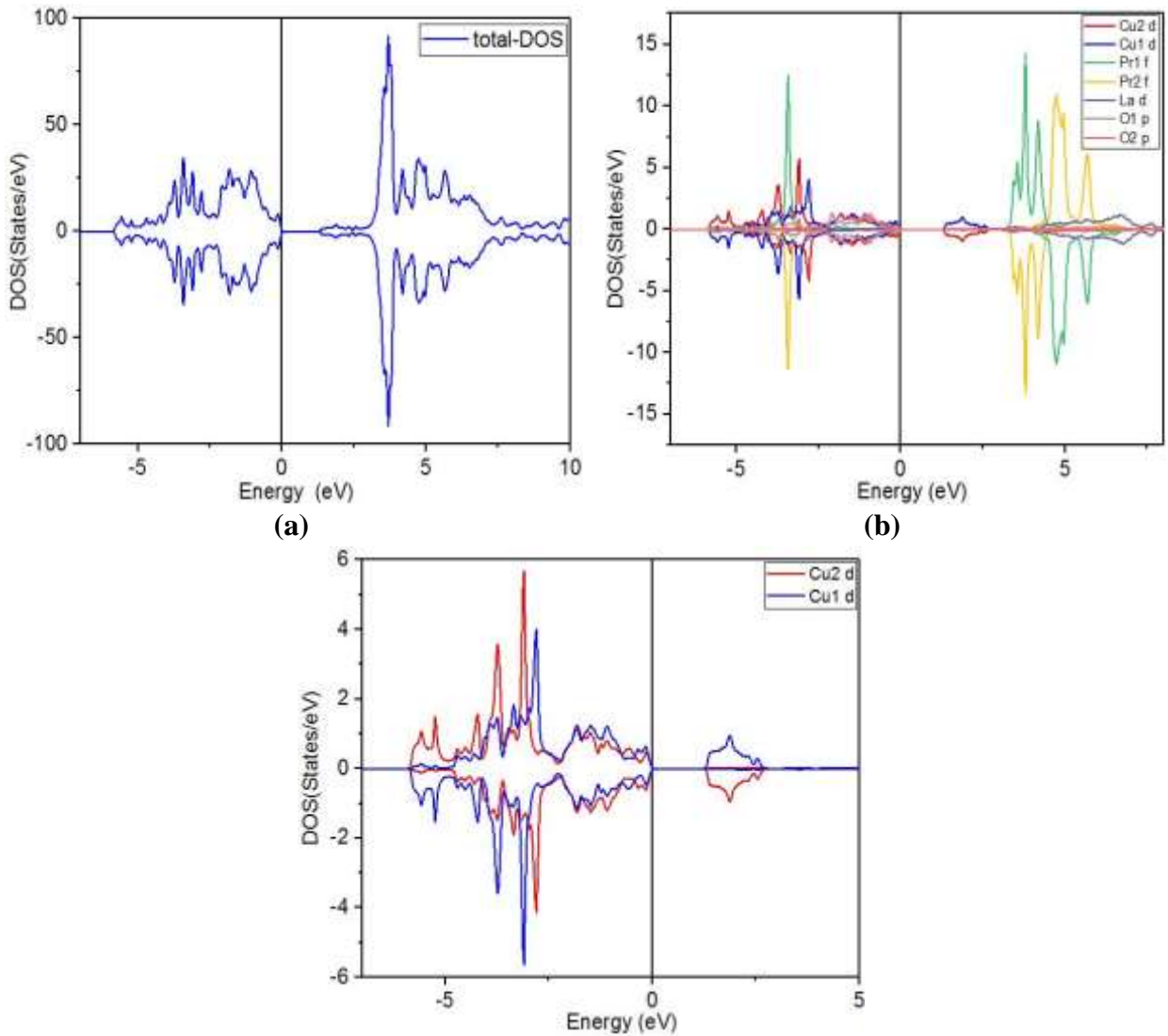

**(c)**

**Figure 9.** (a) Total DOS for configuration A of {La}{Pr}CuO$_4$ in GGA+$U$ approximation, (b) Partial DOS for configuration A of {La}{Pr}CuO$_4$, and (c) Partial DOS of spin-up and spin-down, Cu1 and Cu2.

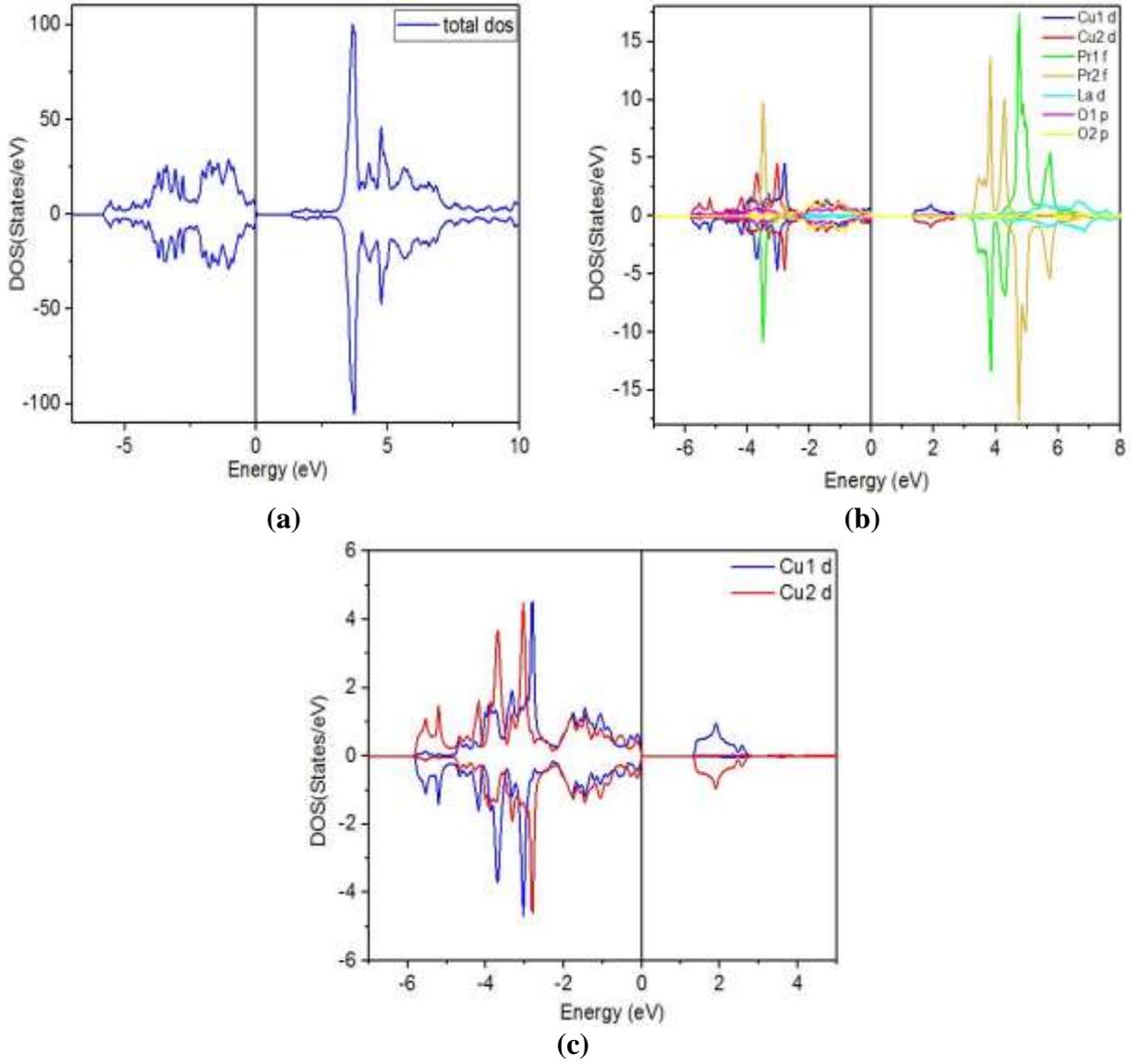

**(a)** **(b)**

**(c)**

**Figure 10.** (a) Total DOS for configuration B of {La}{Pr}CuO$_4$ in GGA+$U$ approximation, (b) Partial DOS for configuration B, and (c) Partial DOS of spin-up and spin-down, Cu1 and Cu2.

**Table 5.** The partial magnetic moments, total magnetic moment, and gap of {La}{Pr}CuO$_4$.

|  |  |  | Configuration A | Configuration B |
|---|---|---|---|---|
| Approximation |  |  | GGA+U | GGA+U |
| Magnetic moment (μ$_B$) | La |  | 0.0015 | 0.0014 |
|  | Cu | Cu1 | 0.6 | 0.6 |
|  |  | Cu2 | -0.6 | -0.6 |
|  | Pr | Pr1 | 1.99 | 1.99 |
|  |  | Pr2 | -1.99 | -1.99 |
| Total magnetic moment (μ$_B$) |  |  | 0 | 0 |

| | | |
|---|---|---|
| Gap (eV) | 1.330 | 1.352 |

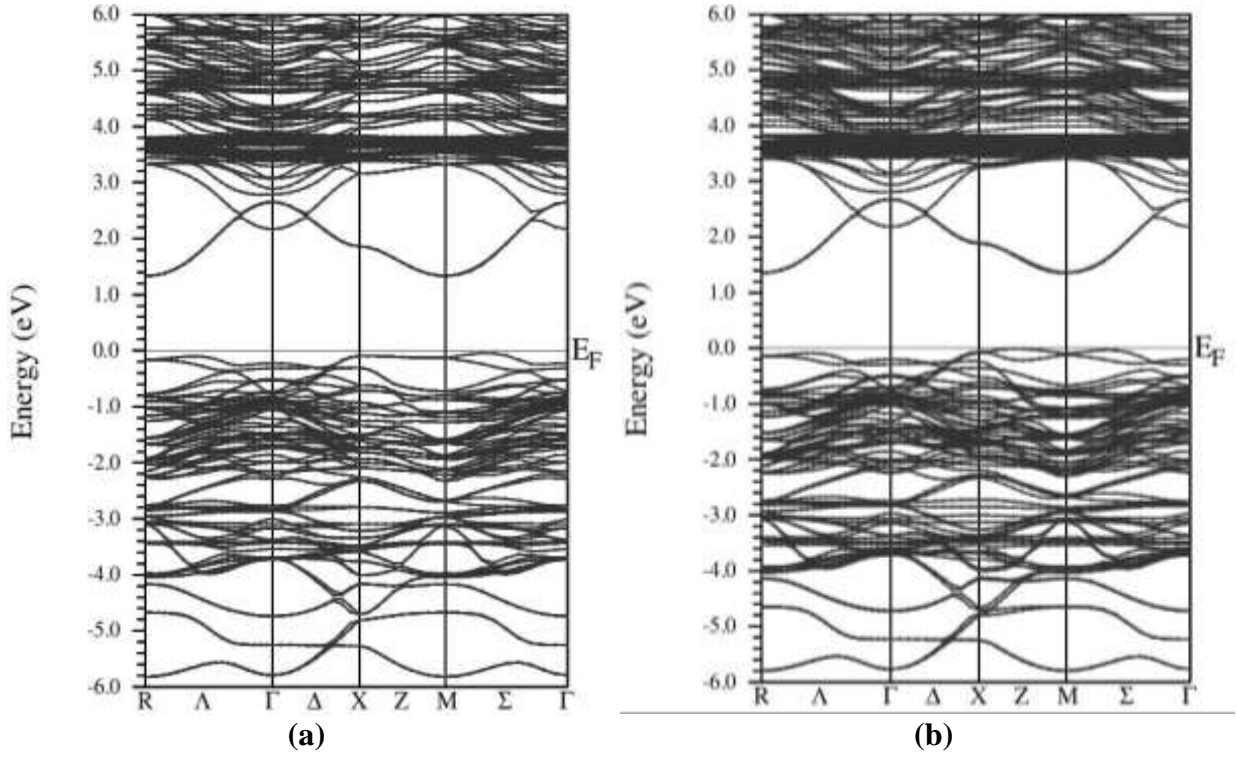

**Figure 11.** Band structure of {La}{Pr}CuO$_4$ for (a) configuration A, and (b) configuration B.

### 2.3. Spatial charge density for {La}{Pr}CuO$_4$

In Figure 12, we show the total charge density in the (001) and (100) planes for both configurations A and B. Because of the similarity between the results obtained for up and down spins, we only report on the results of the up spins. From Figures 12a and 12c, we observe that there is a polar covalent bond between copper and oxygen. This bond is a result of the overlap between the d-Cu and O-p orbitals. The Pr atoms are isolated, suggesting that the bond between the Pr and O atoms is ionic due to the weak overlap between f-Pr and p-O (Figures 12b and 12d). Similarly, the bond between La and O atoms is ionic.

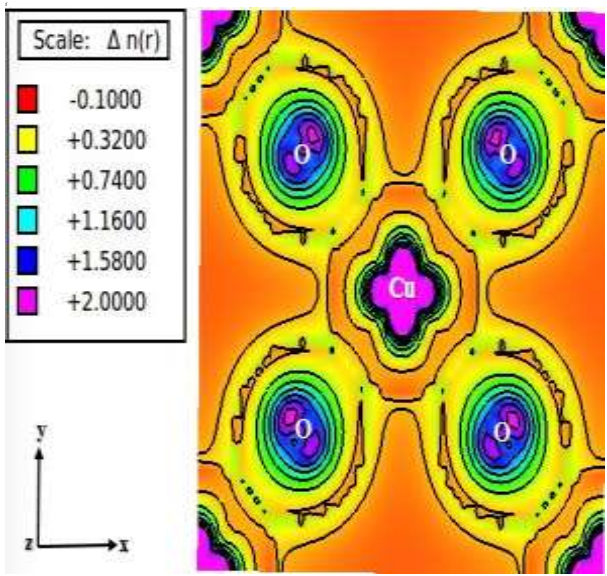
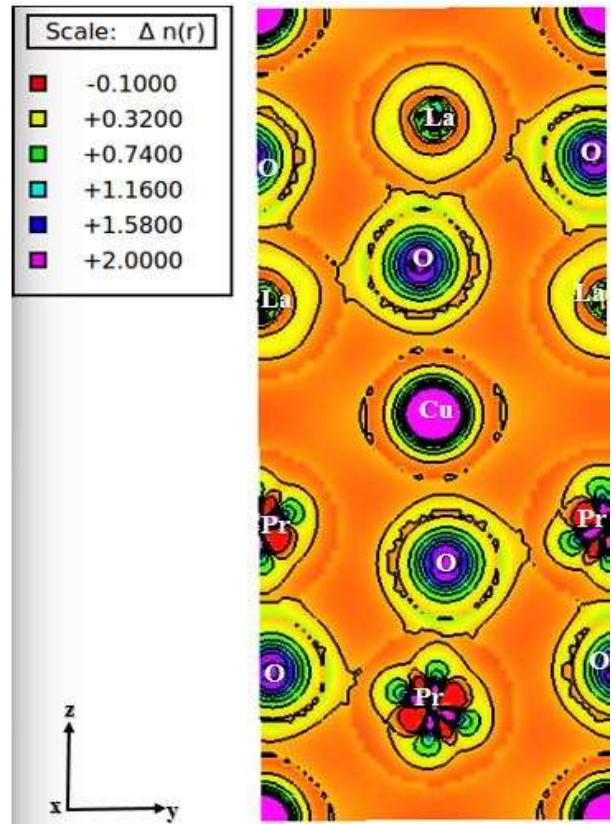

(a)  (b)

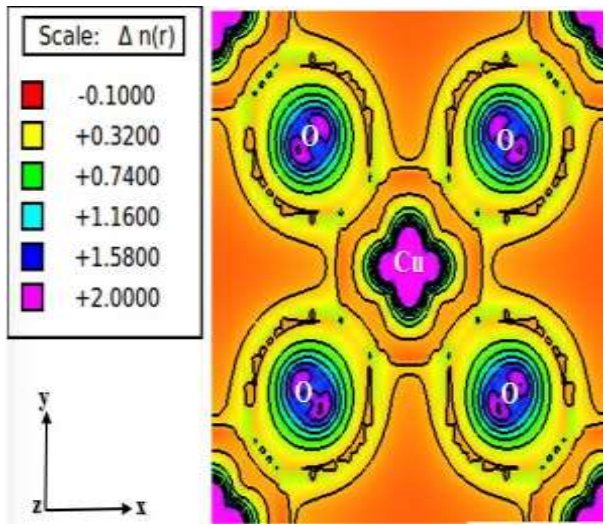
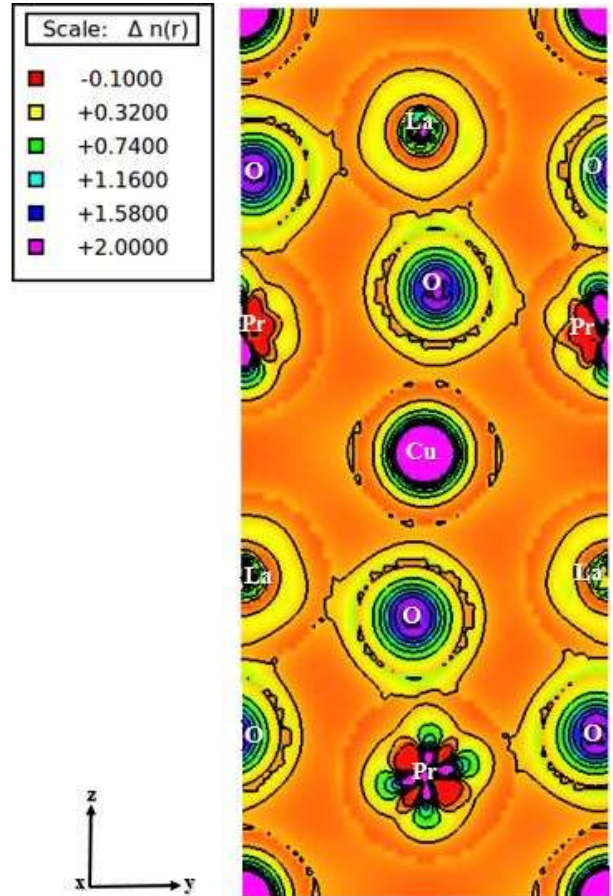

(c)  (d)

**Figure 12.** Charge density of {La}{Pr}CuO$_4$ in configuration A: (a) (001) plane, (b) (100) plane. Charge density of {La}{Pr}CuO$_4$ in configuration B: (c) (001) plane, (d) (100) plane.

## 2.4. Optical conductivity for {La}{Pr}CuO$_4$

Using GGA+$U$ approximation, we calculated the optical conductivities along $x$ ($ab$) and $z$ ($c$) directions for the two configurations of {La}{Pr}CuO$_4$ with energy ranging from 0 to 14 eV. As displayed in Figures 13a and 13b the maximum value for the $ab$ conductivity in configuration A, 6.16×10$^3$ $\Omega^{-1}$cm$^{-1}$, occurs at energy 6.57 eV. For configuration B, the maximum 6.2×10$^3$ $\Omega^{-1}$cm$^{-1}$ occurs at 6.51 eV. Meanwhile, the optical conductivity calculated for La$_2$CuO$_4$ has a maximum value of 7.5×10$^3$ $\Omega^{-1}$cm$^{-1}$ at 8.6 eV. Also, in Figures 13a and 13b, characteristic features at $E_1$ = 1.13 eV, $E_2$ = 1.67 eV, $E_3$ = 1.91 eV, and $E_4$ = 2.42 eV are shown. The energy $E_1$ corresponds to the optical gap. The other energies, $E_2$, $E_3$ and $E_4$, are due to transitions in directions R, X and Γ from the VB to the CB, respectively. The optical gap of {La}{Pr}CuO$_4$ is indirect and its value calculated using Tauc's equation (2) is 1.132 eV for configuration A and 1.138 eV for configuration B (Figures 13c and 13d).

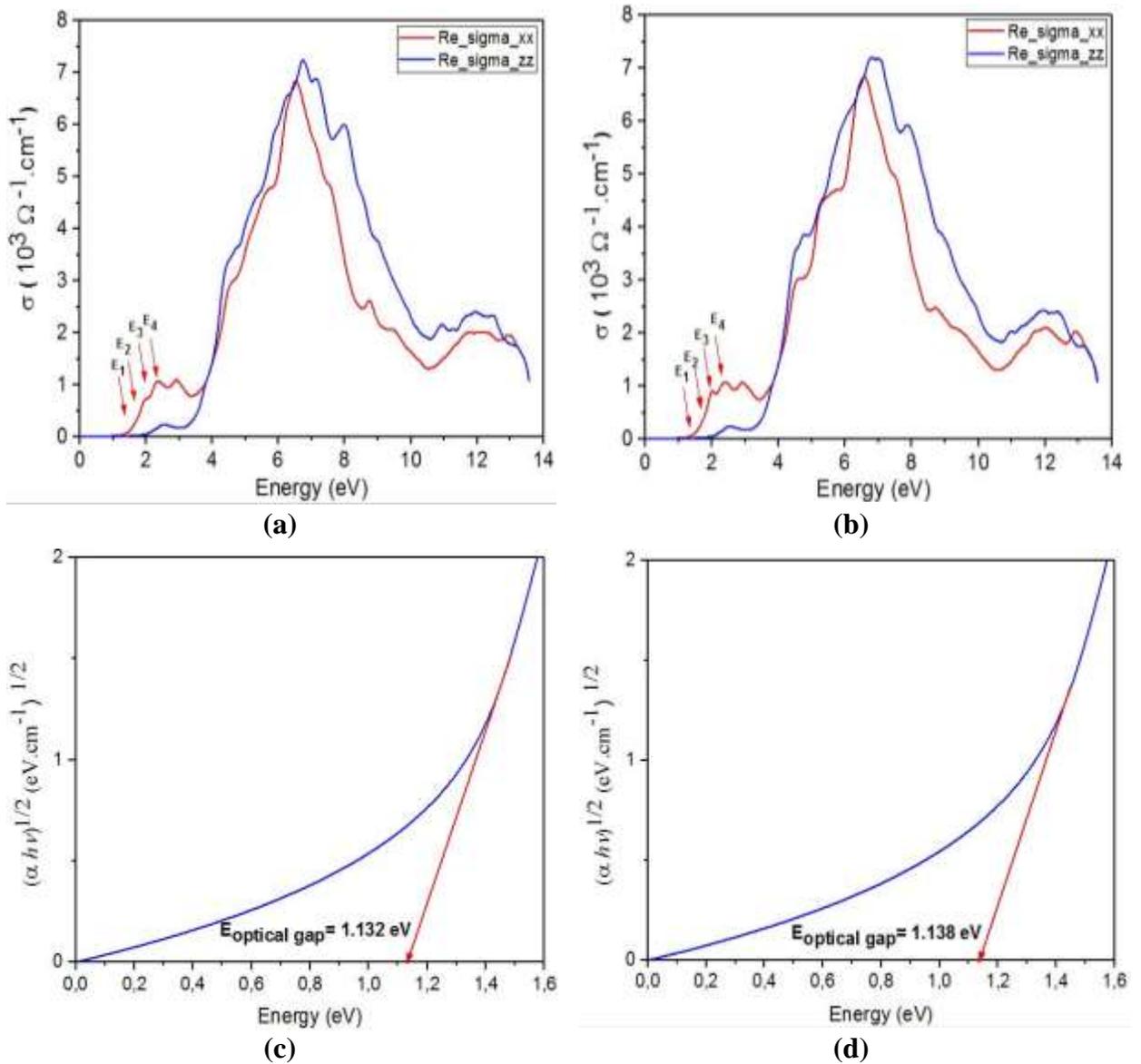

**Figure 13.** Optical conductivity of {La}{Pr}CuO$_4$: (a) Configuration A. (b) Configuration B. Optical gap: (c) Configuration A. (d) Configuration.

## 3. The hypothetical compound {La}{V}CuO$_4$

In our search for a copper-oxide material that could display metallicity in presence of strong AF correlations or even AF order at half filling, we substituted praseodymium, Pr, by vanadium, V. The latter has a smaller size when ionized and can lose up to 5 valence electrons. In principle, in the proposed compound, the valence of V should be only 3+ to satisfy neutrality requirement. Our Wien2k simulations confirm this result. The formation energy of {La}{V}CuO$_4$ calculated for both configurations A and B reveals that this compound is stable, and that configuration B is more stable than configuration A:

$$E_{Formation\,(A)} = E_{system} - E_{La} - E_V - E_{Cu} - 4E_O$$
$$= -67817.5506\ R_y,$$

$$E_{Formation\,(B)} = E_{system} - E_{La} - E_V - E_{Cu} - 4E_O$$
$$= -67817.5688\ R_y.$$

For this new hypothetical material, {La}{V}CuO$_4$, we calculated the XRD profile, and performed DFT calculations of the band structure, DOS, and optical conductivity within the GGA+$U$ approximation.

### 3.1. X-ray profile

The comparison of the lattice parameters of {La}{V}CuO$_4$ (Table 6) and those of La$_2$CuO$_4$ (Table 2) shows a significant difference, due essentially to the big difference in the ionic radii of vanadium and lanthanum. The ionic radii of La and V, 1.160 Å and 0.74 Å [57] respectively, are indeed very different. The difference between {La}{V}CuO$_4$ and La$_2$CuO$_4$ is also shown by the XRD profiles of La$_2$CuO$_4$ (Figure 3) and {La}{V}CuO$_4$ (Figure 14). In contrast, the XRD profile of {La}{Pr}CuO$_4$ (Figure 8) is very similar to that of La$_2$CuO$_4$ (Figure 3), because the ionic radii of lanthanum and praseodymium have comparable values (1.160 Å for La$^{3+}$ and 1.126 Å for Pr$^{3+}$). As seen in Figure 14a for the XRD profile of configurations A for {La}{V}CuO$_4$, the intensity of peak (001) occurs at 2θ ≈ 7°. But in configuration B, Figure 14b, the most intense peak is (200) and occurs at 2θ ≈ 30°. These peaks shift and difference in intensities that exist in the two XRD of both configurations, could be justified, first by the big difference between the radii of La and V, but also by the large difference in the electronegativities of atoms La (with electronegativity 1.1) and V (with 1.63). Note that for the material {La}{Pr}CuO$_4$, La and Pr have comparable radii and comparable electronegativities; 1.1 for La and 1.13 for Pr. This could explain why the XRD profiles of the two configurations are very similar.

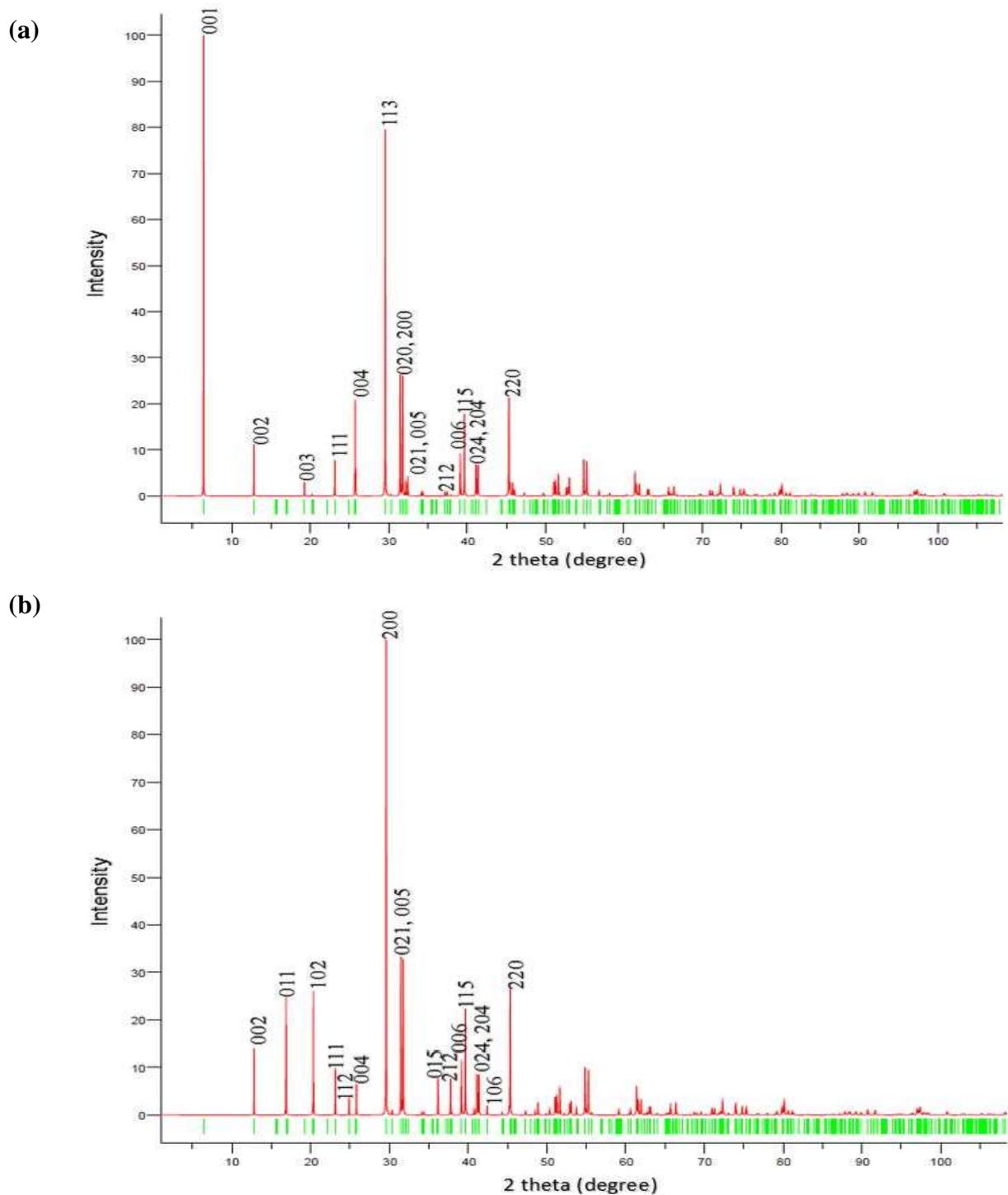

**Figure 14.** The XRD profile calculated by VESTA [29] for (a) {La}{V}CuO$_4$ without mixing of La and V in the charge reservoir, and (b) {La}{V}CuO$_4$ with mixing between La and V.

The parameters of the unit cell $a$, $b$, and $c$, and angles $\alpha$, $\beta$, and $\gamma$ for the two {La}{V}CuO$_4$ structure configurations are shown in Table 6. Notice that the lattice parameters for {La}{V} CuO$_4$ are larger than those of {La}{Pr}CuO$_4$ by a factor 1.4%. This difference can be attributed to the difference in the atomic radii of V and Pr, in the same way are found by T. Kenjo et al. [58], when they replaced by an atom with larger ionic radii the lattice parameters decrease.

**Table 6.** The lattice parameters optimized of {La}{V}CuO$_4$.

| Parameter and angle | $a$ (Å) | $b$ (Å) | $c$ (Å) | $\alpha$ | $\beta$ | $\gamma$ |
|---|---|---|---|---|---|---|
| GGA+U Configuration A | 5.628 | 5.680 | 13.797 | 90° | 90° | 90° |
| GGA+U Configuration B | 5.621 | 5.685 | 13.783 | 90° | 90° | 90° |

### 3.2. Band structure and DOS

Figures 15a and 16a display the total DOS for configurations A and B of {La}{V}CuO$_4$. Clearly, the states in both cases are metallic even with copper's onsite Coulomb repulsion being $U=4$ eV. In the band structure (Figure 17) for the two configurations we observe a small gap above the Fermi energy. Also, in Figures 15b and 16b, partial DOS are shown for La, V, Cu, and O atoms. We observe that the orbital d of the atom V has a higher DOS compared to other atoms. On the other hand, the atom that dominates the DOS near the Fermi level is V, which means that V is thus responsible for the metallic character of the proposed materials. In Figures 15b and 16b, we see that the DOS is dominated by d-Cu and p-O states near energy -3 eV away from the Fermi level, whereas the DOS is dominated by d-La state for energies 3 to 5 eV. Copper's spins are ordered antiferromagnetically as Figures 15c and 16b display. The magnetic moment of V is $2.56\mu_B$ for configuration A and $2.6\mu_B$ for configuration B (Table 7). These values are consistent with $3\mu_B$ for $S = 3/2$ if the Landé factor is $g = 2$, [59].

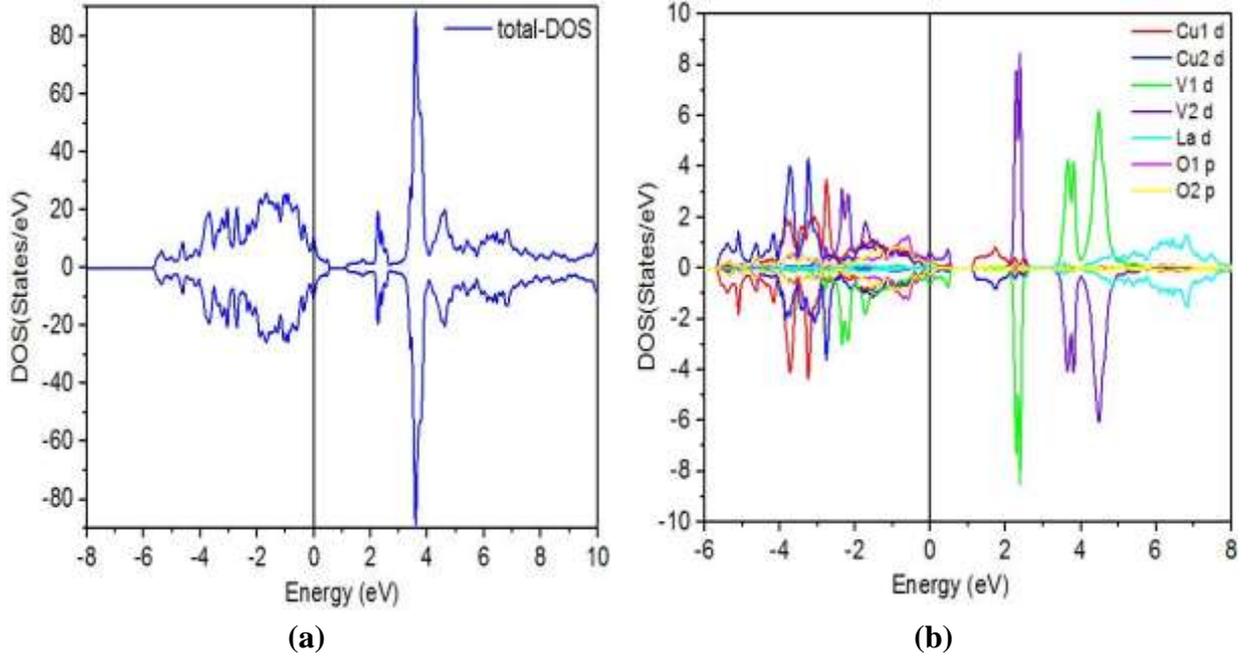

(a)      (b)

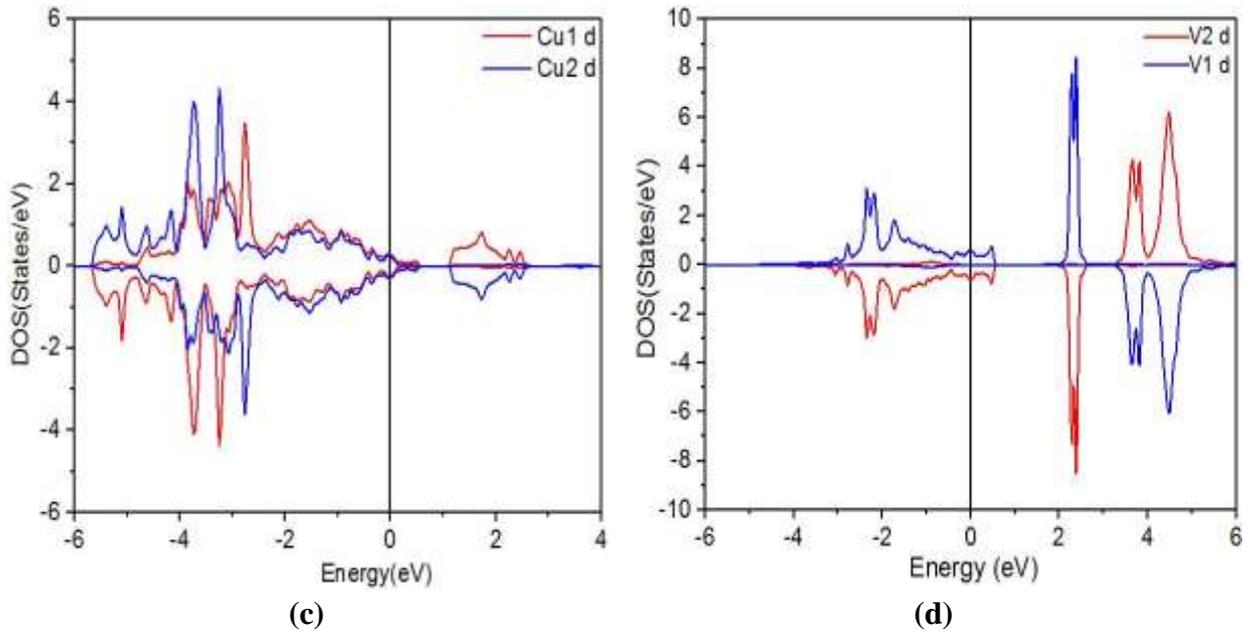

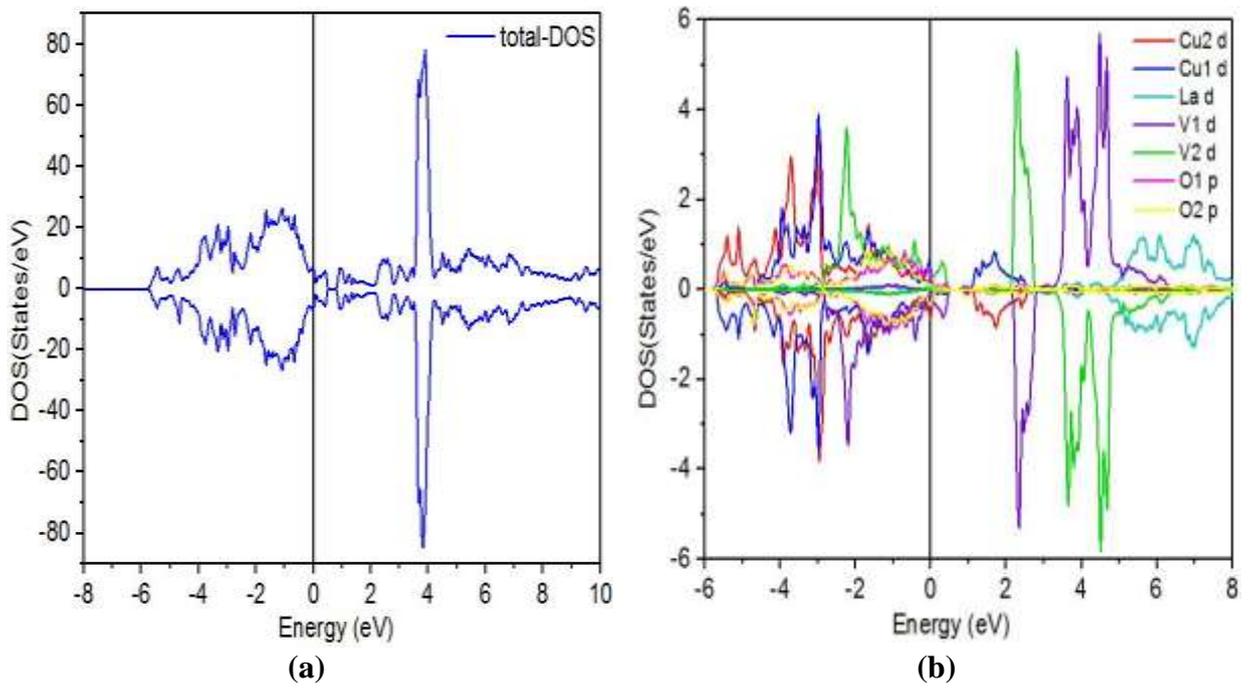

**Figure 15.** (a) Total DOS of configuration A of {La}{V}CuO$_4$ in the GGA+$U$ approximation, (b) Partial DOS of configuration A for {La}{V}CuO$_4$, (c) Partial DOS of spin-up and spin-down Cu1 and Cu2, and (d) Partial DOS of spin-up and spin-down V1 and V2.

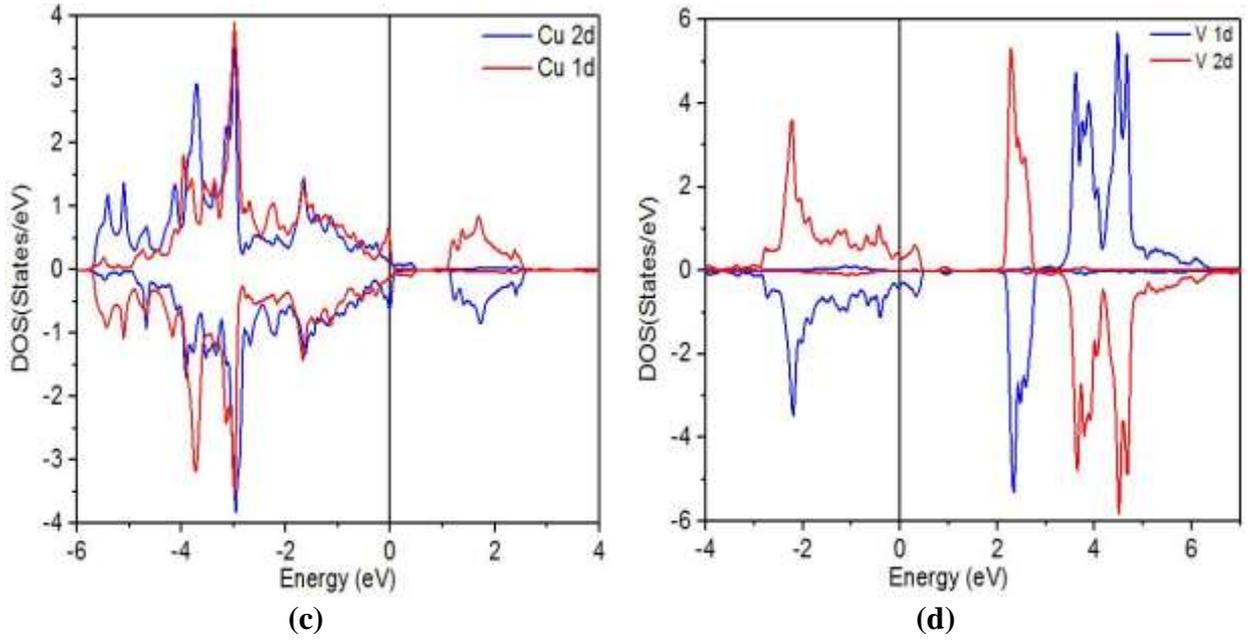

**(c)** **(d)**

**Figure 16.** (a) Total DOS of configuration B for {La}{V}CuO$_4$ in GGA+$U$ approximation, (b) Partial DOS of configuration B, (c) Partial DOS of spin-up and spin-down Cu1 and Cu2, and (d) Partial DOS of spin-up and spin-down V1 and V2.

**Table 7.** The partial magnetic moments, total magnetic moment of {La}{V}CuO$_4$.

|  |  |  | Configuration A | Configuration B |
|---|---|---|---|---|
| Approximation |  |  | GGA+U | GGA+U |
| Magnetic moment ($\mu_B$) |  | La | 0.00035 | 0.00018 |
|  | Cu | Cu1 | 0.59 | 0.6 |
|  |  | Cu2 | -0.59 | -0.6 |
|  | V | V1 | 2.56 | 2.6 |
|  |  | V2 | -2.56 | -2.6 |
| Total magnetic moment ($\mu_B$) |  |  | 0.0468 | 0.0169 |

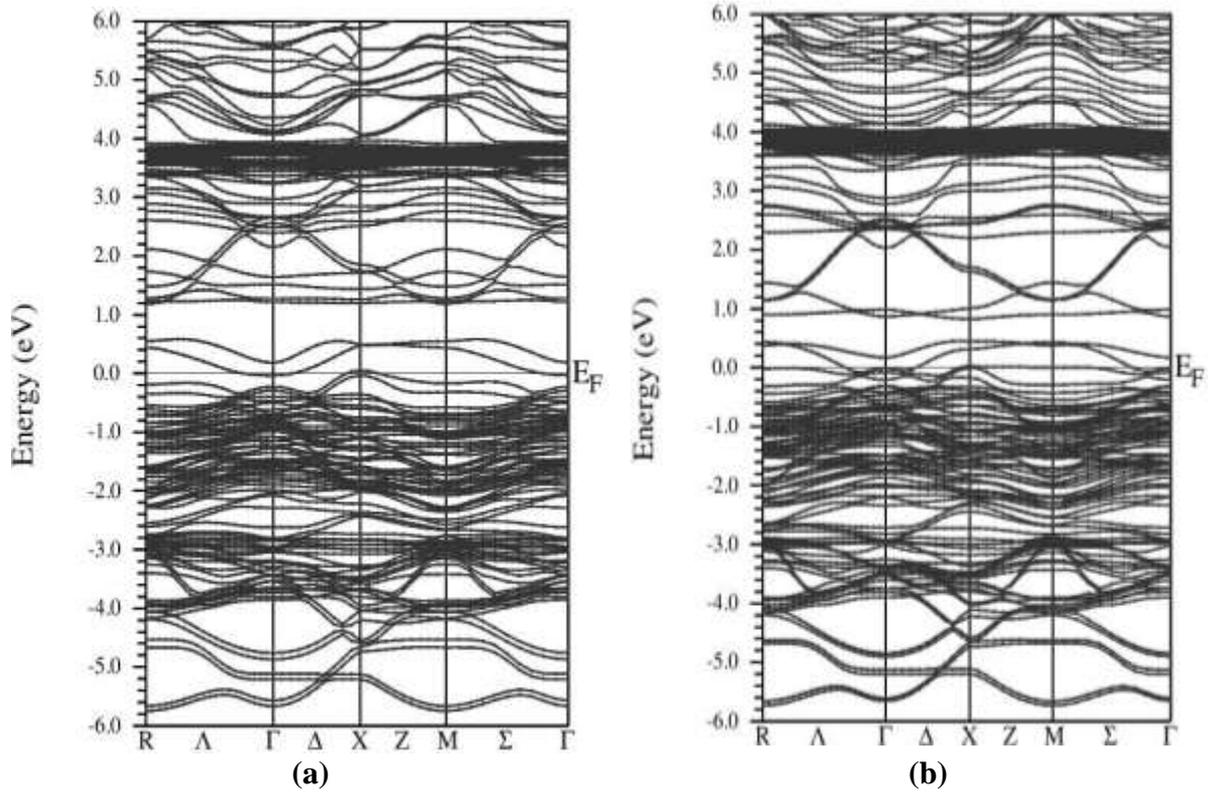

**Figure 17.** Band structure of {La}{V}CuO$_4$ in (a) configuration A, and (b) configuration B.

### 3.3. Spatial charge density for {La}{V}CuO$_4$

The charge density of {La}{V}CuO$_4$ in the (001) and (100) planes for both configurations A and B are shown in Figure 18. A strong polar covalent bond is observed between Cu and O atoms (Figures 18a and 18c). In Figures 18b and 18d, we observe the ionic character of the V-O bond due to the weak interaction of their orbitals (d-V and p-O). We also observe the absence of orbital overlap between La and O, which is consistent with an ionic bond. This agrees with the DOS.

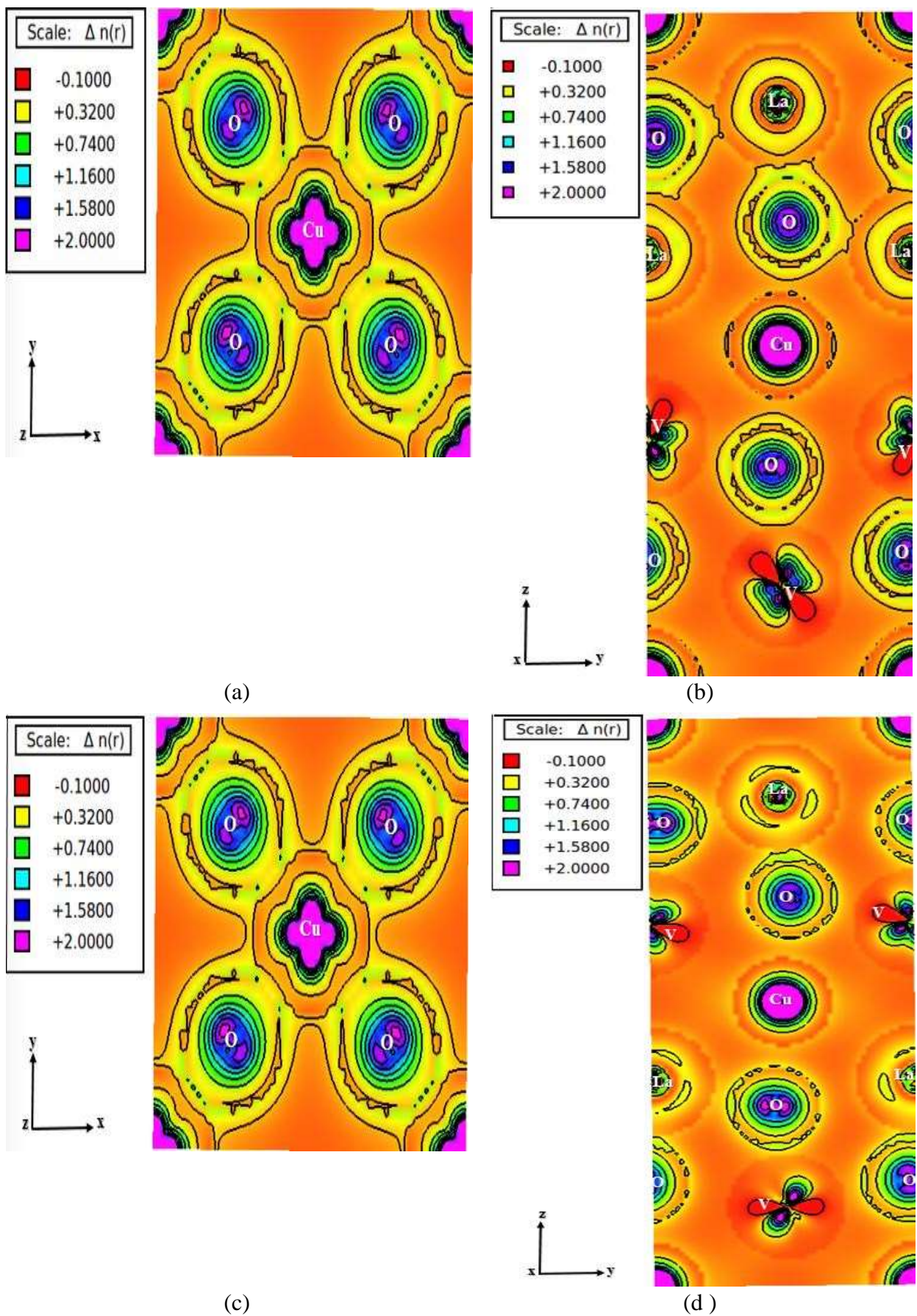

**Figure 18.** Charge density of {La}{V}CuO$_4$ in configuration A: (a) (001) plane, (b) (100) plane. Charge density of {La}{V}CuO$_4$ in configuration B: (c) (001) plane, (d) (100) plane.

### 3.4. Optical conductivity for {La}{V}CuO$_4$

The optical properties calculated by the DFT for {La}{V}CuO$_4$ are displayed in Figure 19. In this proposed material, the conductivity shows no gap near zero energy because this compound is metallic, unlike La$_2$CuO$_4$ and {La}{Pr}CuO$_4$, which are characterized by a gap. The maxima of the optical conductivity for La$_2$CuO$_4$ and {La}{Pr}CuO$_4$ occur at higher energies than those of {La}{V}CuO$_4$. The coexistence of the metallic behavior and AF order at half filling for the V-compound is an important feature. It would be interesting to investigate the dynamics of the charge and spin excitations in this state to figure out what kind of correlated collective behaviors can arise.

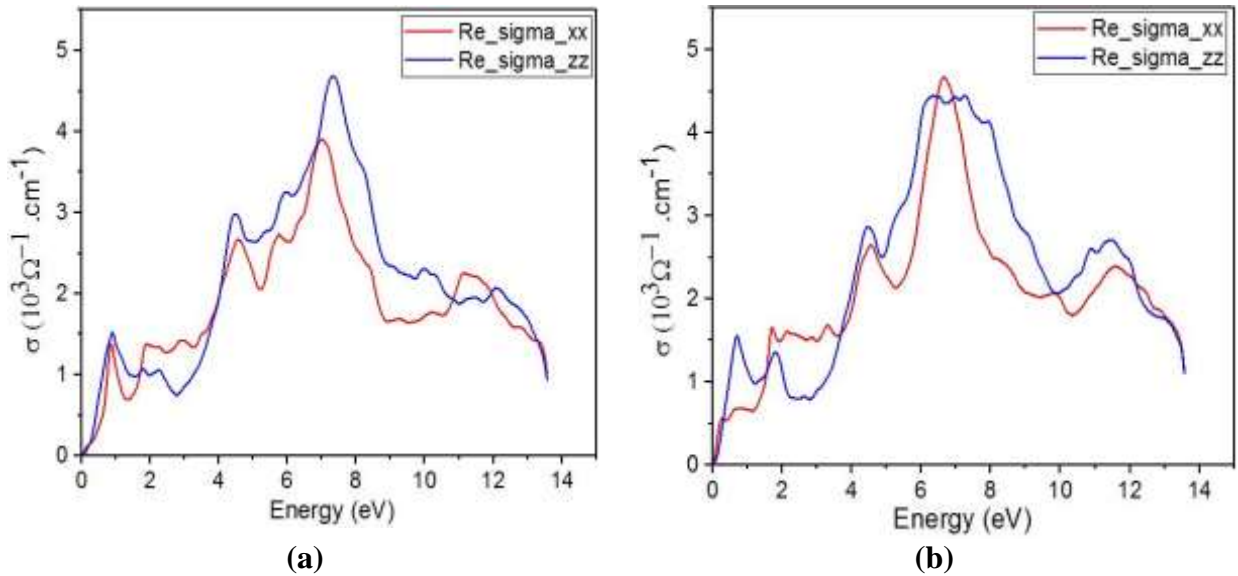

**Figure 19.** Optical conductivity of {La}{V}CuO$_4$ in (a) Configuration A, and (b) Configuration B.

### IV. Conclusions

In our search for exotic materials with ground states mixing strong AF correlations or AF order with metallicity, we investigated the structural, electronic, and optical properties of two hypothetical materials based on copper-oxide high-$T_C$ cuprates. We outlined a new way for growing such materials by depositing, layer after layer, layers of the charge reservoir and CuO$_2$ layers. The positions of the atoms in the charge reservoirs matter. We examined the cases where different atoms mix or not in the same layer of the charge reservoir. Specifically, we report on two hypothetical materials, {La}{Pr}CuO$_4$, which is found to be an AF Mott insulator in both configurations of mixing or non-mixing between La and Pr, and {La}{V}CuO$_4$, which is found to be a quantum metal. We attribute the significant difference in the ground states of these two materials to the significant difference of the radii of atoms Pr and V. Our work should be able to stimulate further studies both experimentally and theoretically. One of the interesting issues that can be addressed is the effect of doping on these hypothetical materials, and the possibility of realizing new states like superconductivity. Also, the gap found in the Mott insulator {La}{Pr}CuO$_4$ is smaller than in La$_2$CuO$_4$, which may suggest potential applications in optoelectronic systems near the visible spectrum.

We calculated the electronic and optical properties of the proposed material {La}{Pr}CuO$_4$ and {La}{V}CuO$_4$ and compared our results with those of the well-known material La$_2$CuO$_4$. Note that we recalculated the properties of the latter and recovered good agreement with literature data. Our work was done using the linear augmented plane waves method at full potential (FP-LAPW) within the GGA+$U$ approximation in the context of the density functional theory. The Wien2k software was used to perform our computations. The La$_2$CuO$_4$ material is known to be an antiferromagnetic insulator due to the strong local Coulomb repulsion. The proposed material {La}{Pr}CuO$_4$ is found to be insulating and antiferromagnetically ordered, but the material {La}{V}CuO$_4$ shows a metallic behavior coexisting with long-range AF order. This state could constitute a true ground state and it is thus termed a quantum metal, in contrast to the conventional metallic states which are not stable against perturbations leading to ordered and gapped states as temperature approaches 0 K. The magnetic moment of Cu in {La}{V}CuO$_4$ is practically 0.6$\mu_B$ and is comparable to the one in La$_2$CuO$_4$. Our results show that the optical conductivity in the *ab* plane and along *z* direction displays no gap in the undoped antiferromagnetic quantum metal {La}{V}CuO$_4$ because of gapless electronic excitations.

**Reference**

[1] K. A. J. G. Bednorz, & Müller, "Possible highT c superconductivity in the Ba− La− Cu− O system," *Zeitschrift für Phys. B Condens. Matter*, vol. 64, no. 5, pp. 189–193, Oct. 1986, doi: 10.1007/BF01303701.

[2] M. K. Wu *et al.*, "Superconductivity at 93 K in a new mixed-phase Yb-Ba-Cu-O compound system at ambient pressure," *Phys. Rev. Lett.*, vol. 58, no. 9, pp. 908–910, 1987, doi: 10.1103/PhysRevLett.58.908.

[3] J. R. Bardeen, J., Cooper, L. N., & Schrieffer, "Theory of super conductivity," *Phys. Rev.*, vol. 108, pp. 1–332, 1957, doi: 10.1103/PhysRevLett.58.908.

[4] A. P. Drozdov, M. I. Eremets, I. A. Troyan, V. Ksenofontov, and S. I. Shylin, "Conventional superconductivity at 203 K at high pressures," *Nature*, vol. 525, pp. 73–76, 2015, doi: 10.1038/nature14964.

[5] M.Yankowitz et al., "Tuning superconductivity in twisted bilayer graphene," *Science (80-. ).*, vol. 363, pp. 1059–1064, 2019, doi: 10.1126/science.aav1910.

[6] T. E. Weller, M. Ellerb, S. S. Saxena, R. P. Smith, and N. T. Skipper, "Superconductivity in the intercalated graphite compounds C6Yb and C6Ca," *Nat. Phys.*, vol. 1, no. 1, pp. 39–41, 2005, doi: 10.1038/nphys0010.

[7] A. P. Drozdov *et al.*, "Superconductivity at 250 K in lanthanum hydride under high pressures," *Nature*, vol. 569, no. 7757, pp. 528–531, 2019, doi: 10.1038/s41586-019-1201-8.

[8] Y. Cao *et al.*, "Unconventional superconductivity in magic-angle graphene superlattices," *Nature*, vol. 556, no. 7699, pp. 43–50, 2018, doi: 10.1038/nature26160.

[9] P. Dai, B. C. Chakoumakos, G. F. Sun, K. W. Wong, and D. F. Lu, "Synthesis and neutron powder diffraction study of the superconductor HgBa2Ca2Cu3O8 + by Tl substitution," vol. 243, pp. 201–206, 1995.

[10] M. Azzouz, "Rotating antiferromagnetism in high-temperature superconductors," no. August 2001, pp. 1–12, 2003, doi: 10.1103/PhysRevB.67.134510.

[11] M. Azzouz, "Thermodynamics of high- T," no. November, pp. 1–12, 2003, doi: 10.1103/PhysRevB.68.174523.

[12] M. Azzouz, "Chemical potentials of high-temperature superconductors Mohamed," vol. 052501, no. August, pp. 4–7, 2004, doi: 10.1103/PhysRevB.70.052501.

[13] H. Saadaoui and M. Azzouz, "Doping dependence of coupling between charge carriers and bosonic modes in the normal state of high-," *Phys. Rev. B*, pp. 1–11, 2005, doi: 10.1103/PhysRevB.72.184518.

[14] M. Azzouz and B. W. Ramakko, "The electronic structure of the high- T C cuprates within the hidden rotating order," *J. Phys. Condens. MATTER*, vol. 345605, 2010, doi: 10.1088/0953-8984/22/34/345605.

[15] M. Azzouz, "Fermi Surface Reconstruction due to Hidden Rotating Antiferromagnetism in N and P-Type High-," pp. 215–232, 2013, doi: 10.3390/sym5020215.

[16] L. D. M. Peter Blaha, Karlheinz Schwarz, Georg K. H. Madsen, Dieter Kvasnicka, Joachim Luitz, Robert Laskowski, Fabien Tran, *An Augmented Plane Wave Plus Local Orbitals Program for Calculating Crystal Properties*, vol. 2. 2001.

[17] G. K. H. Madsen, P. Blaha, K. Schwarz, and E. Sjo, "Efficient linearization of the augmented plane-wave method ¨," vol. 64, pp. 1–9, 2001, doi: 10.1103/PhysRevB.64.195134.

[18] P. HOHENBERG, "Inhomogeneous Electron Gas," *Phys. Rev.*, vol. 136, no. 4, pp. 391–402, 1964, doi: 10.1103/PhysRev. 136. B864.

[19] R. Brendel and D. Bormann, "An infrared dielectric function model for amorphous solids," *J. Appl. Phys.*, vol. 71, pp. 1–7, 1992, doi: 10.1063/1.350737.

[20] J. P. Perdew, "Accurate Density Functional for the Energy: Real-Space Cutoff of the Gradient Expansion for the Exchange Hole," *Phys. Rev. Lett.*, vol. 55, no. 16, pp. 1665–1668, 1985, doi: 10.1103/PhysRevLett.55.1665.

[21] P. Perdew, "Generalized correlation : gradient approximations for exchange and A look backward and forward," *Phys. Rev. B - Condens. Matter Mater. Phys.*, vol. 172, no. 2, 1991, doi: 10.1016/0921-4526(91)90409-8.

[22] J. P. Perdew, K. Burke, and M. Ernzerhof, "Generalized Gradient Approximation Made Simple," *Phys. Rev. Lett.*, no. 3, pp. 3865–3868, 1996, doi: 10.1103/PhysRevLett.77.3865.

[23] V. I. Anisimov, J. Zaanen, and O. K. Andersen, "Band theory and Mott insulators: Hubbard U instead of Stoner I," *Phys. Rev. B*, vol. 44, no. 3, 1991, doi: 10.1103/PhysRevB.44.943.

[24] B. Himmetoglu, A. Floris, S. De Gironcoli, and M. Cococcioni, "Hubbard-Corrected DFT Energy Functionals : The LDA 1 U Description of Correlated Systems," vol. 1151738, 2013, doi: 10.1002/qua.24521.

[25] S. Pesant and M. Cˆ, "DFT + U study of magnetic order in doped La 2 CuO 4 crystals," *Phys. Rev. B*, vol. 085104, pp. 1–7, 2011, doi: 10.1103/PhysRevB.84.085104.

[26] L. M. Kolchina *et al.*, "RSC Advances fuel cells †," *RSC Adv.*, vol. 6, pp. 101029–101037, 2016, doi: 10.1039/C6RA21970E.

[27] F. J. Twagirayezu, "Density functional theory study of the effect of Vanadium doping on electronic and optical properties of NiO," *Int. J. Comput. Mater. Sci. Eng.*, vol. 8, no. 2, pp. 1–13, 2019, doi: 10.1142/S2047684119500076.

[28] A. Kokalj, "XCrySDen-a new program for displaying crystalline structures and electron densities," *J. Mol. Graph. Model.*, vol. 17, no. 3–4, pp. 176–179, 1999, doi: 10.1016/S1093-3263(99)00028-5.

[29] K. Momma and F. Izumi, "VESTA 3 for three-dimensional visualization of crystal , volumetric and morphology data," pp. 1272–1276, 2011, doi: 10.1107/S0021889811038970.

[30] M. Reehuis *et al.*, "Crystal structure and high-field magnetism of La2 Cu O4," *Phys. Rev. B - Condens. Matter Mater. Phys.*, vol. 73, no. 14, pp. 1–8, 2006, doi: 10.1103/PhysRevB.73.144513.

[31] M. T. Czyyk and G. A. Sawatzky, "Local-density functional and on-site correlations: The electronic structure of La2CuO4 and LaCuO3," *Phys. Rev. B*, vol. 49, no. 20, pp. 14211–14228, 1994, doi: 10.1103/PhysRevB.49.14211.

[32] F. Birch, "Finite elastic strain of cubic crystals," *Phys. Rev.*, vol. 71, no. 11, pp. 809–824,

1947, doi: 10.1103/PhysRev.71.809.

[33] J. D. Jorgensen *et al.*, "Superconducting phase of La2CuO4+: A superconducting composition resulting from phase separation," *Phys. Rev. B*, vol. 38, no. 16, pp. 11337–11345, 1988, doi: 10.1103/PhysRevB.38.11337.

[34] J. W. Furness *et al.*, "An accurate first-principles treatment of doping-dependent electronic structure of high-temperature cuprate superconductors," *Commun. Phys.*, vol. 1, no. 1, pp. 1–6, 2018, doi: 10.1038/s42005-018-0009-4.

[35] M. A. Nillasari and B. Kurniawan, "Structural and morphological characterization of cuprate superconductor La2-xSrxCuO4 (x=0 and x=0.07) synthesized by sol-gel method," *J. Phys. Conf. Ser.*, vol. 1170, no. 1, pp. 0–5, 2019, doi: 10.1088/1742-6596/1170/1/012060.

[36] Z. X. Shen, W. E. Spicer, D. M. King, D. S. Dessau, and B. O. Wells, "Photoemission studies of high-Tc superconductors: The superconducting gap," *Science (80-. ).*, vol. 267, no. 5196, pp. 343–350, 1995, doi: 10.1126/science.267.5196.343.

[37] I. Santoso *et al.*, "Unraveling local spin polarization of Zhang-Rice singlet in lightly hole-doped cuprates using high-energy optical conductivity," *Phys. Rev. B*, vol. 95, no. 16, pp. 1–13, 2017, doi: 10.1103/PhysRevB.95.165108.

[38] I. J. Punkkinen, M. P. J., Kokko, K., Hergert, W., & Väyrynen, "Fe2O3 within the LSDA + U approach," *J. Phys. Condens. MATTER*, vol. 2341, 1999, doi: 10.1088/0953-8984/11/11/006.

[39] J. K. Perry, J. Tahir-Kheli, and W. A. Goddard, "Antiferromagnetic band structure of La2CuO4: Becke-3-Lee-Yang-Parr calculations," *Phys. Rev. B - Condens. Matter Mater. Phys.*, vol. 63, no. 14, pp. 1–6, 2001, doi: 10.1103/PhysRevB.63.144510.

[40] C. Lane *et al.*, "Antiferromagnetic Ground State of La 2 CuO 4 : A Parameter-free Ab Initio Description," *Phys. Rev. B*, pp. 1–6, 2018, doi: 10.1103/PhysRevB.98.125140.

[41] H. E. Vaknin, D., Sinha, S. K., Moncton, D. E., Johnston, D. C., Newsam, J. M., Safinya, C. R., & King Jr, "Antiferromagnetism in La2Cu04-y," *Phys. Rev. Lett.*, vol. 58, no. 26, pp. 2802–2805, 1987, doi: 10.1103/PhysRevLett.58.2802.

[42] S. N. Storchak, V. G., Brewer, J. H., Eshchenko, D. G., Mengyan, P. W., Parfenov, O. E., Tokmachev, A. M., ... & Barilo, "Local magnetic order in La 2 CuO 4 seen via μ + SR spectroscopy Local magnetic order in La 2 CuO 4 seen via μ + SR spectroscopy," *J. Phys. Conf. Ser.*, 2014, doi: 10.1088/1742-6596/551/1/012024.

[43] A. J. Ginder, J. M., Roe, M. G., Song, Y., McCall, R. P., Gaines, J. R., Ehrenfreund, E., & Epstein, "Photoexcitations in La 2CuO4: 2-eV energy gap and long-lived defect states," *Phys. Rev. B*, vol. 37, no. 13, pp. 0–3, 1988, doi: 10.1103/PhysRevB.37.7506.

[44] J. P. Sun, J., Ruzsinszky, A., & Perdew, "Strongly constrained and appropriately normed semilocal density functiona," *Phys. Rev. Lett.*, 2015, doi: 10.1103/PhysRevLett.115.036402.

[45] A. Chakraborty, M. Dixit, D. Aurbach, and D. T. Major, "Predicting accurate cathode properties of layered oxide materials using the SCAN meta-GGA density functional," *npj Comput. Mater.*, vol. 4, no. 1, pp. 46–49, 2018, doi: 10.1038/s41524-018-0117-4.

[46] D. I. Badrtdinov *et al.*, "Hybridization and spin-orbit coupling effects in the quasi-one-dimensional," *Phys. Rev. B*, vol. 054435, pp. 1–11, 2016, doi: 10.1103/PhysRevB.94.054435.

[47] M. Khenata, R., Bouhemadou, A., Sahnoun, M., Reshak, A. H., Baltache, H., & Rabah, "Elastic , electronic and optical properties of ZnS , ZnSe and ZnTe under pressure Elastic

[47] , electronic and optical properties of ZnS , ZnSe and ZnTe under pressure," *Comput. Mater. Sci.*, no. January, 2006, doi: 10.1016/j.commatsci.2006.01.013.

[48] S. Uchida, S., Ido, T., Takagi, H., Arima, T., Tokura, Y., & Tajima, "Optical spectra ofLaz Sr„Cuo4. Effect of carrier doping on the electronic structure ofthe Cu02 plane S.," *Phys. Rev. B*, vol. 43, no. 10, 1991, doi: 10.1103/PhysRevB.43.7942.

[49] M.Alouani, O. Jepsen, & O. K. Andersen, "Interband optical conductivity of La2CuO4," *Phys. C Supercond.*, vol. 155, pp. 1233–1234, 1988, doi: 10.1016/0921-4534(88)90257-2.

[50] E. J. Baerends, O. V. Gritsenko, and R. Van Meer, "The Kohn-Sham gap, the fundamental gap and the optical gap: The physical meaning of occupied and virtual Kohn-Sham orbital energies," *Phys. Chem. Chem. Phys.*, vol. 15, no. 39, pp. 16408–16425, 2013, doi: 10.1039/c3cp52547c.

[51] L. Hedin, "New method for calculating the one-particle Green's function with application to the electron-gas problem," *Phys. Rev.*, vol. 139, no. 3A, 1965, doi: 10.1103/PhysRev.139.A796.

[52] F. Sottile, M. Marsili, V. Olevano, and L. Reining, "Efficient ab initio calculations of bound and continuum excitons in the absorption spectra of semiconductors and insulators," *Phys. Rev. B - Condens. Matter Mater. Phys.*, vol. 76, no. 16, pp. 1–4, 2007, doi: 10.1103/PhysRevB.76.161103.

[53] G. Onida, L. Reining, and A. Rubio, "Electronic excitations: Density-functional versus many-body Green's-function approaches," *Rev. Mod. Phys.*, vol. 74, no. 2, pp. 601–659, 2002, doi: 10.1103/RevModPhys.74.601.

[54] J. Tauc et al., "Optical properties and electronic structure of amorphous germanium," *Phys. Stat. Sol.*, vol. 15. pp. 627–637, 1966.

[55] R. Kuberkar, D. G., Rajarajan, A. K., Balakrishnan, G., Gupta, L. C., & Vijayaraghavan, "Structural and superconducting properties of the system La1− xPrSrxCuO4," *Phys. C Supercond.*, vol. 182, pp. 149–152, 1991, doi: 10.1016/0921-4534(91)90472-B.

[56] A. Tsukada et al., "New class of T′-structure cuprate superconductors," *Solid State Commun.*, vol. 133, no. 7, pp. 427–431, 2005, doi: 10.1016/j.ssc.2004.12.011.

[57] R. M. Vol, Z. Shengkui, Y. Z. N. Zhoulanl, W. Zhixing, and C. Qiyuan, "Synthesis and electrochemical properties of Al-doped LiVPOZ cathode materials for lithium-ion batteries," vol. 26, no. 5, pp. 445–449, 2007, doi: 10.1016/S1001-0521(07)60243-5.

[58] T. Kenjo and S. Yajima, " Semiconducting Properties of (Ln I , Ln II )CuO 4 and of (Ln, A) 2 CuO 4 (Ln=rare earth, A=alkaline earth) ," *Bulletin of the Chemical Society of Japan*, vol. 50, no. 11. pp. 2847–2850, 1977, doi: 10.1246/bcsj.50.2847.

[59] O. Hamidane and A. Meddour, "First-Principle Predictions of Electronic Properties and Half- Metallic Ferromagnetism in Vanadium-Doped Rock-Salt CaS," pp. 6–9, 2019, doi: 10.1007/s11664-019-07087-9.